\documentclass[a4paper,reprint,superscriptaddress,amsmath,amssymb,aps,pra,longbibliography,nofootinbib]{revtex4-1}

\pdfoutput=1

\usepackage[utf8]{inputenc}
\usepackage[T1]{fontenc}
\usepackage[english]{babel}
\usepackage{microtype}
\usepackage{fnpct}

\usepackage{silence}              
\usepackage{tikz}

\makeatletter
\DeclareRobustCommand{\rvdots}{%
  \vbox{
    \baselineskip4\p@\lineskiplimit\z@
    \kern-\p@
    \vspace{1pt}
    \hbox{.}\hbox{.}\hbox{.}
  }}
\makeatother

\usetikzlibrary{quantikz}
\usepackage{pgfplots}
\pgfplotsset{compat=newest}
\usepgfplotslibrary{groupplots}
\pgfdeclareplotmark{caretleft}
{%
  \pgfsetroundjoin
  \pgfsetcornersarced{\pgfpoint{.6pt}{.6pt}}
  \pgfsetlinewidth{.075\pgfplotmarksize}
  \pgfpathmoveto{\pgfqpoint{-.5pt}{0pt}}%
  \pgfpathlineto{\pgfqpoint{-.5\pgfplotmarksize - .5pt}{.5\pgfplotmarksize}}%
  \pgfpathlineto{\pgfqpoint{-.5\pgfplotmarksize - .5pt}{-.5\pgfplotmarksize}}%
  \pgfpathclose
  \pgfusepath{fill,stroke}
}%
\pgfdeclareplotmark{caretright}
{%
  \pgfsetroundjoin
  \pgfsetcornersarced{\pgfpoint{.6pt}{.6pt}}
  \pgfsetlinewidth{.075\pgfplotmarksize}
  \pgfpathmoveto{\pgfqpoint{.5pt}{0pt}}%
  \pgfpathlineto{\pgfqpoint{.5\pgfplotmarksize + .5pt}{.5\pgfplotmarksize}}%
  \pgfpathlineto{\pgfqpoint{.5\pgfplotmarksize + .5pt}{-.5\pgfplotmarksize}}%
  \pgfpathclose
  \pgfusepath{fill,stroke}
}%
\pgfdeclareplotmark{caretleftright}
{%
  \pgfsetroundjoin
  \pgfsetcornersarced{\pgfpoint{.6pt}{.6pt}}
  \pgfsetlinewidth{.075\pgfplotmarksize}
  \pgfpathmoveto{\pgfqpoint{.5pt}{0pt}}%
  \pgfpathlineto{\pgfqpoint{.5\pgfplotmarksize + .5pt}{.5\pgfplotmarksize}}%
  \pgfpathlineto{\pgfqpoint{.5\pgfplotmarksize + .5pt}{-.5\pgfplotmarksize}}%
  \pgfpathclose
  \pgfusepath{fill,stroke}
  \pgfpathmoveto{\pgfqpoint{-.5pt}{0pt}}%
  \pgfpathlineto{\pgfqpoint{-.5\pgfplotmarksize - .5pt}{.5\pgfplotmarksize}}%
  \pgfpathlineto{\pgfqpoint{-.5\pgfplotmarksize - .5pt}{-.5\pgfplotmarksize}}%
  \pgfpathclose
  \pgfusepath{fill,stroke}
}%
\definecolor{darkgray176}{RGB}{176,176,176}
\definecolor{steelblue39125161}{RGB}{39,125,161}
\definecolor{tomato2496568}{RGB}{249,65,68}

\usepackage{mathtools}
\usepackage{physics}
\usepackage{dsfont}
\newcommand{\mat}[1]{\mathbf{#1}}
\renewcommand{\vec}[1]{\underline{#1}}

\usepackage{tabularx}
\usepackage{booktabs}
\usepackage{multirow}

\usepackage[hidelinks]{hyperref}
\usepackage[capitalise]{cleveref}
\setcitestyle{numbers,sort&compress}
\let\doi\undefined                
\usepackage{doi}                  

\usepackage{setspace}             
\usepackage[plain]{algorithm}     
\usepackage[noend]{algpseudocode} 
\algnewcommand\myAnd{\textbf{and} }  %
\algnewcommand\Or{\textbf{or} }
\algnewcommand{\LineComment}[1]{\State \(\triangleright\) #1}
\definecolor{light-gray}{gray}{0.8}
\definecolor{light-light-gray}{gray}{0.95}
\usepackage{etoolbox}
\usetikzlibrary{tikzmark}
\usetikzlibrary{calc}

\errorcontextlines\maxdimen

\newcommand{\ALGtikzmarkcolor}{light-gray}
\newcommand{\ALGtikzmarkextraindent}{5pt}
\newcommand{\ALGtikzmarkverticaloffsetstart}{-1.3ex}
\newcommand{\ALGtikzmarkverticaloffsetend}{-.4ex}
\makeatletter
\newcounter{ALG@tikzmark@tempcnta}

\newcommand\ALG@tikzmark@start{%
    \global\let\ALG@tikzmark@last\ALG@tikzmark@starttext%
    \expandafter\edef\csname ALG@tikzmark@\theALG@nested\endcsname{\theALG@tikzmark@tempcnta}%
    \tikzmark{ALG@tikzmark@start@\csname ALG@tikzmark@\theALG@nested\endcsname}%
    \addtocounter{ALG@tikzmark@tempcnta}{1}%
}

\def\ALG@tikzmark@starttext{start}
\newcommand\ALG@tikzmark@end{%
    \ifx\ALG@tikzmark@last\ALG@tikzmark@starttext
    \else
        \tikzmark{ALG@tikzmark@end@\csname ALG@tikzmark@\theALG@nested\endcsname}%
        \tikz[overlay,remember picture,line width=.75pt, cap=round] \draw[\ALGtikzmarkcolor] let \p{S}=($(pic cs:ALG@tikzmark@start@\csname ALG@tikzmark@\theALG@nested\endcsname)+(\ALGtikzmarkextraindent,\ALGtikzmarkverticaloffsetstart)$), \p{E}=($(pic cs:ALG@tikzmark@end@\csname ALG@tikzmark@\theALG@nested\endcsname)+(\ALGtikzmarkextraindent,\ALGtikzmarkverticaloffsetend)$) in (\x{S},\y{S})--(\x{S},\y{E});%
    \fi
    \gdef\ALG@tikzmark@last{end}%
}

\apptocmd{\ALG@beginblock}{\ALG@tikzmark@start}{}{\errmessage{failed to patch}}
\pretocmd{\ALG@endblock}{\ALG@tikzmark@end}{}{\errmessage{failed to patch}}
\makeatother

\begin{document}

\title{Exploring \textit{ab initio} machine synthesis of quantum circuits}

\author{Richard Meister}
\affiliation{Department of Materials, University of Oxford, Oxford OX1 3PH, United Kingdom}
\author{Cica Gustiani}
\affiliation{Department of Materials, University of Oxford, Oxford OX1 3PH, United Kingdom}
\author{Simon C. Benjamin}
\affiliation{Department of Materials, University of Oxford, Oxford OX1 3PH, United Kingdom}
\affiliation{Quantum Motion, 9 Sterling Way, London N7 9HJ, United Kingdom}
\date{\today}

\begin{abstract}
Gate-level quantum circuits are often derived manually from higher level algorithms. While this suffices for small implementations and demonstrations, ultimately automatic circuit design will be required to realise complex algorithms using hardware-specific operations and connectivity.
Here we explore methods for the \textit{ab initio} creation of circuits within a machine, either a classical computer or a hybrid quantum-classical device. We consider a range of techniques including: methods for introducing new gate structures, optimisation of parameterised circuits and choices of cost functions, and efficient removal of low-value gates exploiting the quantum geometric tensor and other heuristics. Using these principles we tackle the tasks of automatic encoding of unitary processes and translation (recompilation) of a circuit from one form to another. Using emulated quantum computers with various noise-free gate sets we provide simple examples involving up to 10 qubits, corresponding to 20 qubits in the augmented space we use. Further applications of specific relevance to chemistry modelling are considered in a sister paper, `Exploiting subspace constraints and ab initio variational methods for quantum chemistry'.

The emulation environments used were \texttt{QuEST}, \texttt{QuESTlink} and \texttt{pyQuEST}. All resources will be made openly accessible and are currently available upon request.
\end{abstract}

\maketitle

\section{Introduction}
The standard formalism for describing instructions on quantum computers is that of circuits consisting of quantum gates \cite{toffoli1981combinatorial,fredkin1982logic,feynman1986quantum,deutsch1989networks}. This description of reversible logic functions serves as a platform on which the capabilities of a given quantum hardware can be described as sets of available \emph{native} gates, as well as a language in which to describe quantum algorithms using well-established universal sets of gates \cite{barenco1995elementary}.

Native gate sets for quantum hardware often consist of a limited number of gates, which express hardware constraints like restricted qubit connectivity or a specific set of supported gates, as is the case e.g. for trapped ions~\cite{akerman2015ions,shapira2018ions,webb2018ions,manovitz2021ions,shuo2021ions}, superconducting qubits~\cite{chow2012superconducting,zhu2021superconducting,long2021superconducting,reuer2021superconducting}, and silicon-based hardware~\cite{wu2018silicon,ferraro2022silicon,evans2022silicon,noiri2022silicon,mills2021silicon}. Quantum algorithms, on the other hand, are often prescribed in a gate set that fits the properties of the computation~\cite{quantalgzoo}. This mismatch in expressibility necessitates methods for \emph{circuit compilation}, where the description in one gate set is translated to an equivalent circuit using a different set, or, more generally, \emph{circuit synthesis}, where a circuit is constructed from a generic description of its unitary action. This process is analogous to the compilation process on classical hardware, where -- before execution -- high-level programming language instructions must be translated to assembly and ultimately to byte code, compatible with the specific hardware it will be executed on.

Several techniques have been proposed to approach such quantum compilation tasks. \emph{Exact unitary decompositions} apply transformations to the target unitary, which yields a solvable relationship between matrix elements of the unitary and gates in the resulting circuit~\cite{tucci1998rudimentary,iten2016isometries,iten2019universalq,krol2022decomp}. These methods always lead to exact -- but sometimes impractically long -- expressions of unitaries as gate sequences. Small problems acting on only a few qubits can also be solved using exhaustive search of all possible gate sequences~\cite{gustiani2021grover} to find the optimal circuit implementing the desired unitary.

In some cases, a known circuit for the desired quantum algorithm may be incompatible with the connectivity constraints of the hardware. This task is often called \emph{qubit routing} or \emph{circuit transformation}, and several methods have been applied to efficiently find the necessary \textsc{swap} operations between gates, including simulated annealing~\cite{zhou2020annealing}, tabu search~\cite{jiang2021tabu}, artificial neural networks~\cite{paler2020machine,zhou2021learning}, and specialised heuristics~\cite{childs2019transformations,li2019tackling,niu2020heuristic,zhou2022tree}. Furthermore, sometimes \textsc{swap}s can be avoided altogether~\cite{devulapalli2022routing}.

A different approach is to dynamically construct a circuit through machine learning, heuristics, or metaheuristics. Several such techniques have been proposed, including pseudorandom walks~\cite{matteo2016parallelizing}, genetic algorithms~\cite{arufe2022genetic}, temporal planning~\cite{venturelli2018temporal}, and deep learning~\cite{moro2021deep}. Such methods can sometimes be used to synthesise approximations to a desired unitary, rather than a perfect re-expression, resulting in shorter circuits than exact compilation. Of particular relevance for the present work are approaches which randomly propose circuit structure changes to try and minimise some cost function. Examples of this are given in the references~\cite{cincio2018overlap,bilkis2021vans,khatri2019quantumassisted}. Each of these works proposes the use of different cost functions, but all of them employ Metropolis-Hastings random sampling to minimise the cost.

It is also noteworthy that while many variational quantum eigensolvers (VQEs) -- which seek to generate a circuit to prepare the ground state of a Hamiltonian from a given input state -- use fixed circuit structures, there have been advancements in generating gate sequences dynamically, for example ADAPT-VQE~\cite{grimsley2019adapt,tang2021qubitadapt} and the Evolutionary Variational Quantum Eigensolver (EVQE) \cite{rattew2019evqe}. As we will discuss, these approaches can, in principle, also be used for circuit synthesis tasks.

The present manuscript presents a broad exploration of the efficacy of VQA-based methods for circuit synthesis and (re-)compilation. We leverage previously described methods as well as techniques that are, to the authors' best knowledge, novel. To assess the capabilities of each approach we use emulated quantum computers, and document the results of systematic studies across a variety of contexts (gates sets, connectivities, etc.) and problem scales (up to 20 qubits). Based on the trends we observe, we draw conclusions on the prospects and value of the approach. We note that small-scale circuit manipulation is certainly practical and can be of key significance in realising components of larger circuits. However, the direct synthesis of entire circuits at a scale relevant to quantum advantage does remain a challenging task whose feasibility is unclear. 

The rest of this paper is structured as follows. In \cref{sec:unitary_equiv} we describe the cost function we use to quantify the closeness of a proposed circuit to the target unitary, which heavily draws ideas from Refs.~\cite{khatri2019quantumassisted} and \cite{jones2022robust}. \Cref{sec:synthesis} describes the routines and subroutines we use to generate circuits in detail. We then apply these routines to various problems, report the results in \cref{sec:results}, and discuss their implications as well as potential future improvements in \cref{sec:discussion}.

\section{Unitary equivalence via energy minimisation} \label{sec:unitary_equiv}
\subsection{Previous work}
The present manuscript substantially builds upon earlier formalisms introduced in, for example, Refs.~\cite{khatri2019quantumassisted,jones2022robust}. In this section, we briefly recapitulate their methods and motivate our extensions to it.

When synthesising a circuit $\mathcal{C}$ for a given unitary $U$, some measure of how well $\mathcal{C}$ approximates $U$ is necessary to drive an optimisation routine towards (approximate) equivalence of $\mathcal{C}$ and $U$. \citet{jones2022robust} use the energy of an artificial Hamiltonian to provide such a measure for the case where equivalence is wanted only for a single input state $\ket{\psi_0}$. \citet{khatri2019quantumassisted}, on the other hand, provide a method called \emph{Hilbert-Schmidt test}, which uses the average fidelity $\expval{\mathcal{C}^\dagger U}{\psi}$ over Haar-distributed random states $\ket{\psi}$ to take \emph{all} input states into account when measuring the closeness of $U$ and $\mathcal{C}$. In the following, we start from the formalism in~\cite{jones2022robust} and extend it to arrive at a variant of the Hilbert-Schmidt test, which we will use as our cost function.

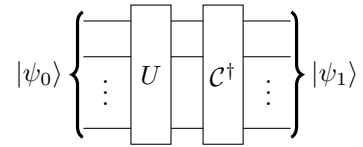
\begin{figure}[tb]
\providecommand{\ket}[1]{\left|#1\right\rangle}
\begin{tikzpicture}[scale=1.000000,x=1pt,y=1pt]
\filldraw[color=white] (0.000000, -6.750000) rectangle (78.000000, 47.250000);
\draw[color=black] (0.000000,40.500000) -- (78.000000,40.500000);
\draw[color=black] (0.000000,27.000000) -- (78.000000,27.000000);
\draw[color=black] (0.000000,0.000000) -- (78.000000,0.000000);
\filldraw[color=white,fill=white] (0.000000,-3.375000) rectangle (-4.000000,43.875000);
\draw[decorate,decoration={brace,amplitude = 4.000000pt},very thick] (0.000000,-3.375000) -- (0.000000,43.875000);
\draw[color=black] (-4.000000,20.250000) node[left] {$\ket{\psi_0}$};
\begin{scope}
\draw (6.000000, 13.500000) node {$\hspace{.5em}\substack{\vspace{-.7em}\\\vdots}$};
\end{scope}
\draw (25.500000,40.500000) -- (25.500000,0.000000);
\begin{scope}
\draw[fill=white] (25.500000, 20.250000) +(-45.000000:10.606602pt and 37.123106pt) -- +(45.000000:10.606602pt and 37.123106pt) -- +(135.000000:10.606602pt and 37.123106pt) -- +(225.000000:10.606602pt and 37.123106pt) -- cycle;
\clip (25.500000, 20.250000) +(-45.000000:10.606602pt and 37.123106pt) -- +(45.000000:10.606602pt and 37.123106pt) -- +(135.000000:10.606602pt and 37.123106pt) -- +(225.000000:10.606602pt and 37.123106pt) -- cycle;
\draw (25.500000, 20.250000) node {$U$};
\end{scope}
\draw (52.500000,40.500000) -- (52.500000,0.000000);
\begin{scope}
\draw[fill=white] (52.500000, 20.250000) +(-45.000000:10.606602pt and 37.123106pt) -- +(45.000000:10.606602pt and 37.123106pt) -- +(135.000000:10.606602pt and 37.123106pt) -- +(225.000000:10.606602pt and 37.123106pt) -- cycle;
\clip (52.500000, 20.250000) +(-45.000000:10.606602pt and 37.123106pt) -- +(45.000000:10.606602pt and 37.123106pt) -- +(135.000000:10.606602pt and 37.123106pt) -- +(225.000000:10.606602pt and 37.123106pt) -- cycle;
\draw (52.500000, 20.250000) node {$\mathcal{C}^\dagger$};
\end{scope}
\begin{scope}
\draw (72.000000, 13.500000) node {$\substack{\vspace{-.7em}\\\vdots}\hspace{.5em}$};
\end{scope}
\filldraw[color=white,fill=white] (78.000000,-3.375000) rectangle (82.000000,43.875000);
\draw[decorate,decoration={brace,mirror,amplitude = 4.000000pt},very thick] (78.000000,-3.375000) -- (78.000000,43.875000);
\draw[color=black] (82.000000,20.250000) node[right] {$\ket{\psi_1}$};
\end{tikzpicture}
    \caption{The setup used in Ref.~\cite{jones2022robust} to synthesise a circuit $\mathcal{C}$ which has the same action on $\ket{\psi_0}$ as $U$. The target is $\ket{\psi_1} \sim \ket{\psi_0}$.}
    \label{fig:state_recompile}
\end{figure}

The technique in~\cite{jones2022robust} starts by first applying the target unitary $U$ to the target input state $\ket{\psi_0}$. Then, the task is to find a circuit $\mathcal{C}^\dagger$ that inverts the action of $U$, such that at the output $\ket{\psi_1}\coloneqq\mathcal{C}^\dagger U\ket{\psi_0}$ the initial state $\ket{\psi_0}$ is recovered up to a global phase\footnote{In this work we use $\sim$ between operators to denote equivalence up to a global phase, i.e. $A \sim B \leftrightarrow A = e^{i\theta} B$.}. If the output is proportional to the input, $\ket{\psi_0} \sim \ket{\psi_1}$, then necessarily $\mathcal{C} \sim U$ holds for the input state $\ket{\psi_0}$. Using an appropriately constructed gapped Hamiltonian $\tilde{H}$ whose ground state is $\ket{\psi_0}$, the initial problem is now an energy minimisation task -- as depicted in \cref{fig:state_recompile} -- of the form
\begin{equation} \label{eq:state_recompile}
    \min_{\mathcal{C}} \ev{U^\dagger \mathcal{C} \tilde{H} \mathcal{C}^\dagger U}{\psi_0}.
\end{equation}

However, it is often desirable to synthesise a circuit $\mathcal{C}$ that completely recovers the action of a given unitary $U$ for \emph{all} relevant input states. This might be the full Hilbert space of $\ket{\psi}$ -- we call it $\mathcal{H}$ -- or a closed subspace thereof, depending on the application. Recent research \cite{bilkis2021vans,cincio2021machine,caro2021learning,caro2022learning,gibbs2022learning} has shown that in some cases, using a subset of $k \ll \dim(\mathcal{H})$ random states $\ket{\psi_k}$ sampled from $\mathcal{H}$, and minimising the sum of their energies
\begin{equation}
    \min_{\mathcal{C}} \sum_k \ev{U^\dagger \mathcal{C} \tilde{H} \mathcal{C}^\dagger U}{\psi_k}
\end{equation}
is a sufficient condition to get unitary equivalence. However, this only holds true for highly structured operators. Therefore, in this work, we use cost functions building on Ref.~\cite{khatri2019quantumassisted} to probe all states in the Hilbert space simultaneously, rather than restricting to a random sample thereof.

\begin{figure}[tb]
\providecommand{\ket}[1]{\left|#1\right\rangle}
\begin{tikzpicture}[scale=1.000000,x=1pt,y=1pt]
\filldraw[color=white] (0.000000, -6.750000) rectangle (178.000000, 101.250000);
\draw[color=black] (0.000000,94.500000) -- (178.000000,94.500000);
\draw[color=black] (0.000000,81.000000) -- (178.000000,81.000000);
\draw[color=black] (0.000000,54.000000) -- (178.000000,54.000000);
\filldraw[color=white,fill=white] (0.000000,50.625000) rectangle (-4.000000,97.875000);
\draw[decorate,decoration={brace,amplitude = 4.000000pt},very thick] (0.000000,50.625000) -- (0.000000,97.875000);
\draw[color=black] (-4.000000,74.250000) node[left] {$\ket{0}_\mathcal{H}$};
\draw[color=black] (0.000000,40.500000) -- (178.000000,40.500000);
\draw[color=black] (0.000000,27.000000) -- (178.000000,27.000000);
\draw[color=black] (0.000000,0.000000) -- (178.000000,0.000000);
\filldraw[color=white,fill=white] (0.000000,-3.375000) rectangle (-4.000000,43.875000);
\draw[decorate,decoration={brace,amplitude = 4.000000pt},very thick] (0.000000,-3.375000) -- (0.000000,43.875000);
\draw[color=black] (-4.000000,20.250000) node[left] {$\ket{0}_\mathcal{H'}$};
\begin{scope}
\draw[fill=white] (10.000000, 94.500000) +(-45.000000:8.485281pt and 8.485281pt) -- +(45.000000:8.485281pt and 8.485281pt) -- +(135.000000:8.485281pt and 8.485281pt) -- +(225.000000:8.485281pt and 8.485281pt) -- cycle;
\clip (10.000000, 94.500000) +(-45.000000:8.485281pt and 8.485281pt) -- +(45.000000:8.485281pt and 8.485281pt) -- +(135.000000:8.485281pt and 8.485281pt) -- +(225.000000:8.485281pt and 8.485281pt) -- cycle;
\draw (10.000000, 94.500000) node {$H$};
\end{scope}
\begin{scope}
\draw[fill=white] (10.000000, 81.000000) +(-45.000000:8.485281pt and 8.485281pt) -- +(45.000000:8.485281pt and 8.485281pt) -- +(135.000000:8.485281pt and 8.485281pt) -- +(225.000000:8.485281pt and 8.485281pt) -- cycle;
\clip (10.000000, 81.000000) +(-45.000000:8.485281pt and 8.485281pt) -- +(45.000000:8.485281pt and 8.485281pt) -- +(135.000000:8.485281pt and 8.485281pt) -- +(225.000000:8.485281pt and 8.485281pt) -- cycle;
\draw (10.000000, 81.000000) node {$H$};
\end{scope}
\begin{scope}
\draw[fill=white] (10.000000, 54.000000) +(-45.000000:8.485281pt and 8.485281pt) -- +(45.000000:8.485281pt and 8.485281pt) -- +(135.000000:8.485281pt and 8.485281pt) -- +(225.000000:8.485281pt and 8.485281pt) -- cycle;
\clip (10.000000, 54.000000) +(-45.000000:8.485281pt and 8.485281pt) -- +(45.000000:8.485281pt and 8.485281pt) -- +(135.000000:8.485281pt and 8.485281pt) -- +(225.000000:8.485281pt and 8.485281pt) -- cycle;
\draw (10.000000, 54.000000) node {$H$};
\end{scope}
\begin{scope}
\draw (10.000000, 67.500000) node {$\rvdots$};
\end{scope}
\begin{scope}
\draw (10.000000, 13.500000) node {$\rvdots$};
\end{scope}
\draw (29.000000,94.500000) -- (29.000000,40.500000);
\begin{scope}
\draw[fill=white] (29.000000, 40.500000) circle(3.000000pt);
\clip (29.000000, 40.500000) circle(3.000000pt);
\draw (26.000000, 40.500000) -- (32.000000, 40.500000);
\draw (29.000000, 37.500000) -- (29.000000, 43.500000);
\end{scope}
\filldraw (29.000000, 94.500000) circle(1.500000pt);
\draw (39.000000,81.000000) -- (39.000000,27.000000);
\begin{scope}
\draw[fill=white] (39.000000, 27.000000) circle(3.000000pt);
\clip (39.000000, 27.000000) circle(3.000000pt);
\draw (36.000000, 27.000000) -- (42.000000, 27.000000);
\draw (39.000000, 24.000000) -- (39.000000, 30.000000);
\end{scope}
\filldraw (39.000000, 81.000000) circle(1.500000pt);
\draw (49.000000,54.000000) -- (49.000000,0.000000);
\begin{scope}
\draw[fill=white] (49.000000, 0.000000) circle(3.000000pt);
\clip (49.000000, 0.000000) circle(3.000000pt);
\draw (46.000000, 0.000000) -- (52.000000, 0.000000);
\draw (49.000000, -3.000000) -- (49.000000, 3.000000);
\end{scope}
\filldraw (49.000000, 54.000000) circle(1.500000pt);
\draw (60.000000, 101.250000) node[text width=144pt,above,text centered] {$\ket{\psi_0'}$};
\draw[dashed] (60.000000,-6.750000) -- (60.000000,101.250000);
\draw (77.500000,94.500000) -- (77.500000,54.000000);
\begin{scope}
\draw[fill=white] (77.500000, 74.250000) +(-45.000000:10.606602pt and 37.123106pt) -- +(45.000000:10.606602pt and 37.123106pt) -- +(135.000000:10.606602pt and 37.123106pt) -- +(225.000000:10.606602pt and 37.123106pt) -- cycle;
\clip (77.500000, 74.250000) +(-45.000000:10.606602pt and 37.123106pt) -- +(45.000000:10.606602pt and 37.123106pt) -- +(135.000000:10.606602pt and 37.123106pt) -- +(225.000000:10.606602pt and 37.123106pt) -- cycle;
\draw (77.500000, 74.250000) node {$U$};
\end{scope}
\draw (100.500000,94.500000) -- (100.500000,54.000000);
\begin{scope}
\draw[fill=white] (100.500000, 74.250000) +(-45.000000:10.606602pt and 37.123106pt) -- +(45.000000:10.606602pt and 37.123106pt) -- +(135.000000:10.606602pt and 37.123106pt) -- +(225.000000:10.606602pt and 37.123106pt) -- cycle;
\clip (100.500000, 74.250000) +(-45.000000:10.606602pt and 37.123106pt) -- +(45.000000:10.606602pt and 37.123106pt) -- +(135.000000:10.606602pt and 37.123106pt) -- +(225.000000:10.606602pt and 37.123106pt) -- cycle;
\draw (100.500000, 74.250000) node {$\mathcal{C}^\dagger$};
\end{scope}
\draw (118.000000, 101.250000) node[text width=144pt,above,text centered] {$\ket{\psi_1'}$};
\draw[dashed] (118.000000,-6.750000) -- (118.000000,101.250000);
\draw (129.000000,54.000000) -- (129.000000,0.000000);
\begin{scope}
\draw[fill=white] (129.000000, 0.000000) circle(3.000000pt);
\clip (129.000000, 0.000000) circle(3.000000pt);
\draw (126.000000, 0.000000) -- (132.000000, 0.000000);
\draw (129.000000, -3.000000) -- (129.000000, 3.000000);
\end{scope}
\filldraw (129.000000, 54.000000) circle(1.500000pt);
\draw (139.000000,81.000000) -- (139.000000,27.000000);
\begin{scope}
\draw[fill=white] (139.000000, 27.000000) circle(3.000000pt);
\clip (139.000000, 27.000000) circle(3.000000pt);
\draw (136.000000, 27.000000) -- (142.000000, 27.000000);
\draw (139.000000, 24.000000) -- (139.000000, 30.000000);
\end{scope}
\filldraw (139.000000, 81.000000) circle(1.500000pt);
\draw (149.000000,94.500000) -- (149.000000,40.500000);
\begin{scope}
\draw[fill=white] (149.000000, 40.500000) circle(3.000000pt);
\clip (149.000000, 40.500000) circle(3.000000pt);
\draw (146.000000, 40.500000) -- (152.000000, 40.500000);
\draw (149.000000, 37.500000) -- (149.000000, 43.500000);
\end{scope}
\filldraw (149.000000, 94.500000) circle(1.500000pt);
\begin{scope}
\draw (168.000000, 67.500000) node {$\rvdots$};
\end{scope}
\begin{scope}
\draw (168.000000, 13.500000) node {$\rvdots$};
\end{scope}
\begin{scope}
\draw[fill=white] (168.000000, 94.500000) +(-45.000000:8.485281pt and 8.485281pt) -- +(45.000000:8.485281pt and 8.485281pt) -- +(135.000000:8.485281pt and 8.485281pt) -- +(225.000000:8.485281pt and 8.485281pt) -- cycle;
\clip (168.000000, 94.500000) +(-45.000000:8.485281pt and 8.485281pt) -- +(45.000000:8.485281pt and 8.485281pt) -- +(135.000000:8.485281pt and 8.485281pt) -- +(225.000000:8.485281pt and 8.485281pt) -- cycle;
\draw (168.000000, 94.500000) node {$H$};
\end{scope}
\begin{scope}
\draw[fill=white] (168.000000, 81.000000) +(-45.000000:8.485281pt and 8.485281pt) -- +(45.000000:8.485281pt and 8.485281pt) -- +(135.000000:8.485281pt and 8.485281pt) -- +(225.000000:8.485281pt and 8.485281pt) -- cycle;
\clip (168.000000, 81.000000) +(-45.000000:8.485281pt and 8.485281pt) -- +(45.000000:8.485281pt and 8.485281pt) -- +(135.000000:8.485281pt and 8.485281pt) -- +(225.000000:8.485281pt and 8.485281pt) -- cycle;
\draw (168.000000, 81.000000) node {$H$};
\end{scope}
\begin{scope}
\draw[fill=white] (168.000000, 54.000000) +(-45.000000:8.485281pt and 8.485281pt) -- +(45.000000:8.485281pt and 8.485281pt) -- +(135.000000:8.485281pt and 8.485281pt) -- +(225.000000:8.485281pt and 8.485281pt) -- cycle;
\clip (168.000000, 54.000000) +(-45.000000:8.485281pt and 8.485281pt) -- +(45.000000:8.485281pt and 8.485281pt) -- +(135.000000:8.485281pt and 8.485281pt) -- +(225.000000:8.485281pt and 8.485281pt) -- cycle;
\draw (168.000000, 54.000000) node {$H$};
\end{scope}
\filldraw[color=white,fill=white] (178.000000,-3.375000) rectangle (182.000000,97.875000);
\draw[decorate,decoration={brace,mirror,amplitude = 4.000000pt},very thick] (178.000000,-3.375000) -- (178.000000,97.875000);
\draw[color=black] (182.000000,47.250000) node[right] {$\ket{\psi_1}$};
\end{tikzpicture}
    \caption{The circuit setup equivalent to a Hilbert-Schmidt test in Ref.~\cite{khatri2019quantumassisted}, which we used to synthesise $\mathcal{C}\sim U$. The target is $\ket{\psi_1} \sim \ket{0}_\mathcal{H} \otimes \ket{0}_{\mathcal{H}'}$.}
    \label{fig:fullspace_setup}
\end{figure}
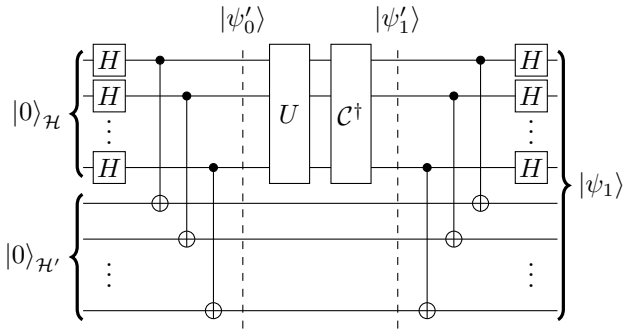

Starting from \cref{eq:state_recompile} and \cref{fig:state_recompile}, it is possible to achieve full unitary equivalence of $U$ and $\mathcal{C}$ by exploiting the Choi--Jamiołkowski isomorphism~\cite{jamil1972linear,choi1975completely}. From an all-zero input we first create the maximally entangled state $\ket{\psi_0'} = \sum_k \ket{k} \otimes \ket{k}$. We can then map the action of $\mathcal{C}^\dagger U$ on every state in $\mathcal{H}$ to the action of $\mathcal{C}^\dagger U \otimes \mathds{1}$ on the single state $\ket{\psi_0'}$ in the space $\mathcal{H} \otimes \mathcal{H'}$, where $\mathcal{H'}$ is a copy of $\mathcal{H}$. \Cref{fig:fullspace_setup} shows a circuit construction of this method, which is equivalent to the Hilbert-Schmidt test introduced in Ref.~\cite{khatri2019quantumassisted}. The output state $\ket{\psi_1'}$ is proportional to the input state $\ket{\psi_0'}$ if and only if $\mathcal{C} \sim U$. Therefore, applying the inverse of the preparation circuit that created $\ket{\psi_0'}$ returns the state to the original computational all-zero state if indeed $\mathcal{C} \sim U$. In order to estimate how close we are to this ideal, we can use any Hamiltonian $\tilde{H}$ which has $\ket{0}_\mathcal{H} \otimes \ket{0}_\mathcal{H'}$ as its unique and gapped ground state as an artificial Hamiltonian. Because we know an expectation value of $\expval*{\tilde{H}}{\psi_1} = 0$ exactly corresponds to $U\sim\mathcal{C}$, energy minimisation techniques can be used to find a circuit $\mathcal{C}$ which is equivalent to a given unitary $U$ for all input states. Note that for $\tilde{H} = \ketbra{0}_\mathcal{H}\otimes\ketbra{0}_{\mathcal{H}'}$, we exactly recover the Hilbert-Schmidt test with a target expectation value of $\expval*{\tilde{H}} = 1$.

\Cref{fig:subspace_setup} in \cref{sec:subspace} shows a -- to the authors' best knowledge -- novel modification to this setup, which interpolates between \cref{fig:state_recompile} and \cref{fig:fullspace_setup} by considering a specific subspace of $\mathcal{H}$, on which we elaborate in \cref{sec:subspace}.

As shorthand notation for our cost function we will write $\expval*{\tilde{H}}$ to mean the expectation value $\expval*{\tilde{H}}{\psi_1}$ in the full augmented space with $\ket{\psi_1}$ as produced by the circuit in \cref{fig:fullspace_setup}. Possible choices for $\tilde{H}$ are discussed in the next subsection.

Using $\expval*{\tilde{H}}$ to determine unitary equivalence ignores global phase factors by which $U$ and $\mathcal{C}$ might differ. In most scenarios, this is desirable, since such a global phase is physically irrelevant. However, if the resulting $\mathcal{C}$ is to be used in a controlled fashion within a larger circuit, a global phase mismatch of $U$ and $\mathcal{C}$ becomes a physically relevant relative phase. In this case, an additional phase gate with an appropriate parameter must be added  to $\mathcal{C}$ at the end of the synthesis process.

\subsection{Synthesis Hamiltonians}\label{sec:cost_functions}
When expressing the condition of unitary equivalence as an energy minimisation problem, there is some freedom in choosing an appropriate $\tilde{H}$, as any gapped Hamiltonian with the all-zero computational basis state as its ground state can be used. In this work, we consider the two Hamiltonians
\begin{equation} \label{eq:H_sum}
    H_\mathrm{sum} = \frac{1}{N}\sum_k \sigma^z_k
\end{equation}
where $\sigma^z_k$ is the Pauli-$Z$ operator acting on qubit $k$, and $N$ is the total number of qubits, and
\begin{equation} \label{eq:H_proj}
    H_\mathrm{proj} = \mathds{1} - \ketbra{0}{0}.
\end{equation}
While $H_\mathrm{proj}$ corresponds -- as previously mentioned -- to the Hilbert-Schmidt test, $H_\mathrm{sum}$ is more closely related to the \emph{local} Hilbert-Schmidt test~\cite{khatri2019quantumassisted}, and the local cost function proposed in~\cite{cerezo2021cost}. Both of these Hamiltonians have a ground state energy of 0 and maximum energy of 1, they differ in the energy structure of their excited states, and thus also in how they measure the closeness of a circuit is to a target unitary.

The projector-based $H_\mathrm{proj}$ simply measures the overlap with the desired all-zero state, while assigning every other product state the same maximum energy of 1.

On the other hand, the sum-based $H_\mathrm{sum}$ assigns each computational basis state an energy proportional to its Hamming distance~\cite{hamming1950error} from the ground state, i.e. it measures how many bit flips would be necessary to get to the desired all-zero state. The state $\ket{0\ldots 01}$ therefore has a lower energy than the state $\ket{1\ldots 11}$ when measured with $H_\mathrm{sum}$, while $H_\mathrm{proj}$ evaluates them as being at equal distances from the target.

The choice of Hamiltonian can also have a significant impact on the difficulty of finding the optimal parameters. For instance, Refs.~\cite{khatri2019quantumassisted,cerezo2021cost}, discuss that global cost functions -- like $H_\mathrm{proj}$ -- are very likely to come across barren plateaus during the parameter optimisation, while employing local costs -- such as $H_\mathrm{sum}$ -- is much less prone to such problems.

Even though in this work we only consider $H_\mathrm{sum}$ and $H_\mathrm{proj}$, many other Hamiltonians may be used. Depending on the application, they can be designed to emphasise different properties of what constitutes a \emph{good} output state, and may penalise some highly undesirable properties more harshly than others.

\subsection{Subspace compilation} \label{sec:subspace}
For some tasks, synthesising the entire unitary is unnecessary. For instance, some chemistry problems demand symmetry conservation, such as particle number preservation and spin preservation. If a unitary with such a structure is applied to a state whose entire weight is within one such subspace, the operation outside that subspace becomes irrelevant. In this task, the unitary $U$ has a block-diagonal form,
\begin{equation}
    U=U_1\oplus U_2\oplus U_3\oplus\dots\oplus U_n,
\end{equation}
where $U_j$ is a unitary and $\oplus$ means the direct sum. The task now is discovering a circuit that correctly implements the unitary $U_j$, given the entire unitary $U$.

We address the compilation within such a subspace as \emph{subspace compilation}. Assuming that the subspace size is much smaller than the full space, we will demonstrate how our learning method considerably improves the synthesis time and decreases the number of gates in the final circuit compared to the compilation in the full space.

Consider a block-diagonal unitary $U$ acting in a Hilbert space $\mathcal H$ of dimension $d=2^n$. Without loss of generality, suppose $U=U_1\oplus U_2$, i.e. a block-diagonal unitary comprised of two blocks. Let $U_1$ be an operator in $\mathcal H_1$ and $U_2$ an operator in $\mathcal H_2$, the unitary $U$ acts on $\mathcal H=\mathcal H_1\oplus\mathcal H_2$, where $\mathcal H_1$ and $\mathcal H_2$ are subspaces of $\mathcal H$. Given that $H_1$ has dimension $d_1$ and $H_2$ has dimension $d_2$, the dimension of $\mathcal H$ is $d_1+d_2=d$.

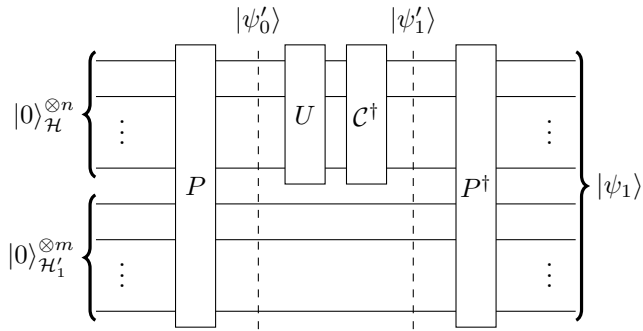
\begin{figure}[tbh]
\providecommand{\ket}[1]{\left|#1\right\rangle}
\begin{tikzpicture}[scale=1.000000,x=1pt,y=1pt]
\filldraw[color=white] (0.000000, -6.750000) rectangle (180.000000, 101.250000);
\draw[color=black] (0.000000,94.500000) -- (180.000000,94.500000);
\draw[color=black] (0.000000,81.000000) -- (180.000000,81.000000);
\draw[color=black] (0.000000,54.000000) -- (180.000000,54.000000);
\filldraw[color=white,fill=white] (0.000000,50.625000) rectangle (-4.000000,97.875000);
\draw[decorate,decoration={brace,amplitude = 4.000000pt},very thick] (0.000000,50.625000) -- (0.000000,97.875000);
\draw[color=black] (-4.000000,74.250000) node[left] {$\ket{0}^{\otimes n}_\mathcal{H}$};
\draw[color=black] (0.000000,40.500000) -- (180.000000,40.500000);
\draw[color=black] (0.000000,27.000000) -- (180.000000,27.000000);
\draw[color=black] (0.000000,0.000000) -- (180.000000,0.000000);
\filldraw[color=white,fill=white] (0.000000,-3.375000) rectangle (-4.000000,43.875000);
\draw[decorate,decoration={brace,amplitude = 4.000000pt},very thick] (0.000000,-3.375000) -- (0.000000,43.875000);
\draw[color=black] (-4.000000,20.250000) node[left] {$\ket{0}^{\otimes m}_{\mathcal{H}'_1}$};
\begin{scope}
\draw (10.000000, 67.500000) node {$\rvdots$};
\end{scope}
\begin{scope}
\draw (10.000000, 13.500000) node {$\rvdots$};
\end{scope}
\draw (37.500000,94.500000) -- (37.500000,0.000000);
\begin{scope}
\draw[fill=white] (37.500000, 47.250000) +(-45.000000:10.606602pt and 75.306872pt) -- +(45.000000:10.606602pt and 75.306872pt) -- +(135.000000:10.606602pt and 75.306872pt) -- +(225.000000:10.606602pt and 75.306872pt) -- cycle;
\clip (37.500000, 47.250000) +(-45.000000:10.606602pt and 75.306872pt) -- +(45.000000:10.606602pt and 75.306872pt) -- +(135.000000:10.606602pt and 75.306872pt) -- +(225.000000:10.606602pt and 75.306872pt) -- cycle;
\draw (37.500000, 47.250000) node {$P$};
\end{scope}
\draw (61.000000, 101.250000) node[text width=144pt,above,text centered] {$\ket{\psi_0'}$};
\draw[dashed] (61.000000,-6.750000) -- (61.000000,101.250000);
\draw (78.500000,94.500000) -- (78.500000,54.000000);
\begin{scope}
\draw[fill=white] (78.500000, 74.250000) +(-45.000000:10.606602pt and 37.123106pt) -- +(45.000000:10.606602pt and 37.123106pt) -- +(135.000000:10.606602pt and 37.123106pt) -- +(225.000000:10.606602pt and 37.123106pt) -- cycle;
\clip (78.500000, 74.250000) +(-45.000000:10.606602pt and 37.123106pt) -- +(45.000000:10.606602pt and 37.123106pt) -- +(135.000000:10.606602pt and 37.123106pt) -- +(225.000000:10.606602pt and 37.123106pt) -- cycle;
\draw (78.500000, 74.250000) node {$U$};
\end{scope}
\draw (101.500000,94.500000) -- (101.500000,54.000000);
\begin{scope}
\draw[fill=white] (101.500000, 74.250000) +(-45.000000:10.606602pt and 37.123106pt) -- +(45.000000:10.606602pt and 37.123106pt) -- +(135.000000:10.606602pt and 37.123106pt) -- +(225.000000:10.606602pt and 37.123106pt) -- cycle;
\clip (101.500000, 74.250000) +(-45.000000:10.606602pt and 37.123106pt) -- +(45.000000:10.606602pt and 37.123106pt) -- +(135.000000:10.606602pt and 37.123106pt) -- +(225.000000:10.606602pt and 37.123106pt) -- cycle;
\draw (101.500000, 74.250000) node {$\mathcal{C}^\dagger$};
\end{scope}
\draw (119.000000, 101.250000) node[text width=144pt,above,text centered] {$\ket{\psi_1'}$};
\draw[dashed] (119.000000,-6.750000) -- (119.000000,101.250000);
\draw (142.500000,94.500000) -- (142.500000,0.000000);
\begin{scope}
\draw[fill=white] (142.500000, 47.250000) +(-45.000000:10.606602pt and 75.306872pt) -- +(45.000000:10.606602pt and 75.306872pt) -- +(135.000000:10.606602pt and 75.306872pt) -- +(225.000000:10.606602pt and 75.306872pt) -- cycle;
\clip (142.500000, 47.250000) +(-45.000000:10.606602pt and 75.306872pt) -- +(45.000000:10.606602pt and 75.306872pt) -- +(135.000000:10.606602pt and 75.306872pt) -- +(225.000000:10.606602pt and 75.306872pt) -- cycle;
\draw (142.500000, 47.250000) node {$P^\dagger$};
\end{scope}
\begin{scope}
\draw (170.000000, 67.500000) node {$\rvdots$};
\end{scope}
\begin{scope}
\draw (170.000000, 13.500000) node {$\rvdots$};
\end{scope}
\filldraw[color=white,fill=white] (180.000000,-3.375000) rectangle (184.000000,97.875000);
\draw[decorate,decoration={brace,mirror,amplitude = 4.000000pt},very thick] (180.000000,-3.375000) -- (180.000000,97.875000);
\draw[color=black] (184.000000,47.250000) node[right] {$\ket{\psi_1}$};
\end{tikzpicture}
\caption{One cost evaluation in a subspace compilation. The unitary $P$ prepares a maximally entangled state $\ket\Phi_{\mathcal{H}\mathcal{H}_1'}$ according to \cref{eq:subspace_phi}.}
\label{fig:subspace_setup}
\end{figure}

Given access to $U$, compiling $U_1$ on subspace $\mathcal H_1$ using our \textit{ab initio} technique has a cost evaluation setup shown in \Cref{fig:subspace_setup}; this circuit generalises the compilation explained in \Cref{sec:cost_functions}.

First, we prepare two quantum registers: the main register (representing $\mathcal{H}$) with $n$ qubits and the ancilla register (representing $\mathcal{H}_1'$) with $m=\lceil\log_2{d_1}\rceil$ qubits. Second, we prepare a maximally entangled state $\ket\Phi_{\mathcal{H}\mathcal{H}_1'}$ on both registers $\mathcal{H} \otimes\mathcal{H}_1'$, where
\begin{equation}\label{eq:subspace_phi}
    P\ket{0}_{\mathcal{H}\mathcal{H}_1'} = \ket{\Phi}_{\mathcal{H}\mathcal{H}_1'}=\frac{1}{\sqrt{d_1}}\sum_{s_j\in S}\ket{s_j}_\mathcal{H}\ket{j}_{\mathcal{H}_1'},
\end{equation}
with a unitary operator $P$, and $S$, an orthonormal basis spanning $\mathcal H_1$.
Note that $\ket\Phi_{\mathcal{H}\mathcal{H}_1'}$ is maximally entangled between spaces $\mathcal{H}$ and $\mathcal{H}_1'$, i.e. $\Tr_\mathcal{H}{(\ketbra\Phi)}_{\mathcal{H}\mathcal{H}_1'} =\mathds{1}_{\mathcal{H}_1'}$ and $\Tr_{\mathcal{H}_1'}{(\ketbra\Phi)}_{\mathcal{H}\mathcal{H}_1'}=\mathds{1}_{\mathcal{H}}$.

\section{Ab initio circuit synthesis} \label{sec:synthesis}

In this section we discuss algorithms and subroutines used to vary the structure of ansatz-circuits and their parameters in order to minimise the expected energy under specific Hamiltonians, thereby solving various circuit synthesis and VQE problems. We introduce notation and definitions for our synthesis protocols momentarily, while \cref{sec:new_gates,sec:param_opt,sec:remove_gates,sec:initial_circ} describe subroutines which are then used within the algorithms laid out in \cref{sec:algorithms}.

\subsection{Formalism}

The building blocks of our circuits are primitive gates $G_k$, which can be single-qubit rotations, multi-qubit rotations, \textsc{swap}s, etc., acting on different sets of qubits. We assume that every gate has a classical parameter, e.g. a rotation angle, associated with it. The formalism is easily extended to also include non-parametrised gates; they are simply gates whose parameter is permanently fixed. In our terminology, altering the parameter associated with a gate does not constitute a replacement of the gate itself.

We refer to a specific set of gates $\mathcal{L} = \{G_k\}$ as a \emph{gate library}. Note that each $G_k$ specifies which qubits the gate acts upon; for example, the Pauli-$X$ rotations on different qubits 1 and 2 -- $R^x_1$ and $R^x_2$ respectively -- would be considered two separate gates $G_k$. This allows a gate library to include information about only locally available gates and qubit connectivity.

A \emph{circuit structure} -- or ansatz -- $\mathcal{C}$ can be represented by an ordered sequence of such primitive gates
\begin{equation}
    \mathcal{C} \coloneqq (\mathcal{C}_0, \mathcal{C}_1, \ldots, \mathcal{C}_{N-1}),
\end{equation}
where $\mathcal{C}_k \in \mathcal{L}$.

Before the circuit can be applied to a quantum state, a parameter vector $\vec{\theta}$ containing the parameter $\theta_k$ for each gate $\mathcal{C}_k$ must be assigned. We write
\begin{equation*}
    \mathcal{C}(\vec{\theta}) \coloneqq (\mathcal{C}_0(\theta_0), \mathcal{C}_1(\theta_1), \ldots, \mathcal{C}_{N-1}(\theta_{N-1})).
\end{equation*}
Applying a circuit $\mathcal{C}(\vec{\theta})$ to a state $\ket{\psi}$ means evaluating
\begin{equation*}
    \mathcal{C}_{N-1}(\theta_{N-1}) \cdot \ldots \cdot \mathcal{C}_1(\theta_{1}) \cdot \mathcal{C}_0(\theta_{0})  \ket{\psi}.
\end{equation*}

A given circuit $\mathcal{C}(\vec{\theta})$ can be modified in two fundamentally different ways. One is to change the parameters $\vec{\theta}$, which we call \emph{parameter optimisation}. The other is to add or remove gates to or from the circuit structure $\mathcal{C}$, which we refer to as \emph{circuit structure modifications}. The procedures for how we perform these modifications are explained in detail in the following subsections. Our routines are also given as simplified versions in high-level pseudocode in \cref{sec:pseudocode}, which might miss some performance enhancing tweaks for the sake of clarity and brevity.

\subsection{Parameter optimisation} \label{sec:param_opt}
For a given circuit structure $\mathcal{C}$ containing parameterised gates -- often called \emph{ansatz circuit} -- we want to find the parameter vector $\vec{\theta}$ which minimises the expected energy of our artificial Hamiltonian $\tilde{H}$, i.e. the cost function. At this minimum, the circuit $\mathcal{C}(\vec{\theta})$ most closely approximates the desired unitary $U$ within the scope of its parameter space. In our algorithms, we use imaginary time evolution~\cite{mcardle2019imaginary,yuan2019variational} in a slightly modified version. We first compute the matrix object\footnote{We use the shorthand notation $\ket{\partial_\mu\psi} \coloneqq \frac{\partial\ket{\psi(\vec{\theta})}}{\partial\theta_{\!\mu}}$.}
\begin{equation} \label{eq:QMT}
    A_{ij} = \Re(\braket{\partial_i \psi}{\partial_j \psi} - \braket{\partial_i \psi}{\psi}\braket{\psi}{\partial_j \psi}),
\end{equation}
which we refer to as the \emph{quantum metric tensor (QMT)} \cite{koczor2019natural,yamamoto2019natural,stokes2020quantumnatural}, and the \emph{gradient vector}
\begin{equation} \label{eq:gradient}
    B_i = -\mel{\partial_i \psi}{H}{\psi}.
\end{equation}
The time evolution of the parameter vector $\vec{\theta}$ is then given by~\cite{yuan2019variational}
\begin{equation}
    \mat{A}\,\dot{\vec{\theta}} = \vec{B}.
\end{equation}
Using the forward Euler method~\cite{euler1769institutionum} we thus get the update rule for the parameter vector
\begin{equation}
    \vec{\theta}_{t+1} = \vec{\theta}_t - \lambda\,\mat{A}^{-1}\,\vec{B}.
\end{equation}
The matrix $\mat{A}$ is often close to singular, so some regularisation method is required to stabilise the iteration. We use Tikhonov regularisation, but other methods may be used as well.

To find a suitable $\lambda$, we use an idea similar to the one presented in Ref.~\cite{jones2022robust}. In each iteration, we start from a small (arbitrary) initial value of $\lambda_0 = 0.05$ for each time step. But, because the evaluation of the QMT and the gradient vector may be expensive operations~\cite{straaten2021cost} compared to the evaluation of the expected energy for a given set of parameters, we then try to exponentially increase the step size until we find a local minimum along the established step direction. This means repeatedly multiplying $\lambda$ by some constant factor $\kappa$ until an energy increase is found, and then accepting the immediately preceding $\lambda$-value. However, if the initial step size turns out to already increase the energy, we instead shrink $\lambda$ exponentially until we find a decrease in energy or hit a minimum step size. This exponential search is almost always useful in emulators, where the gradient direction is known to high numerical precision. Its usefulness in avoiding barren plateaus is reduced on real quantum hardware when shot noise limits how precisely the gradient direction can be determined.

To detect convergence, we use absolute and relative changes of the energy, and require one of the conditions to be met a number $k_\mathrm{conv} \sim 5$ of times. \Cref{algo:min_energy} shows the full procedure.

\subsection{Introduction of new gates} \label{sec:new_gates}

Dynamic expansion of a given ansatz circuit has been explored by VQE methods~\cite{grimsley2019adapt,tang2021qubitadapt,rattew2019evqe} and in more general contexts~\cite{cincio2018overlap,cincio2021machine,bilkis2021vans}. The formalism we introduce here is very generic, but naturally shares some ideas with previous works, especially Ref.~\cite{bilkis2021vans}.

The algorithms introduced later rely on the concept of a \emph{move}, which refers to the most basic possible modification of a circuit structure $\mathcal{C}$, i.e. the insertion of one additional gate at some position in the gate sequence. Given a library of gates $\mathcal{L}$ and a circuit with $N$ gates, it is convenient to define a move as a tuple $(G, n)$, with a gate $G \in \mathcal{L}$ and an index $0 \leq n \leq N$. Applying such a move to an existing circuit $\mathcal{C}$ means inserting gate $G$ at position $n$.
\begin{equation}
    \mathcal{C} \mapsto (\mathcal{C}_0, \ldots, \mathcal{C}_{n - 1}, G, \mathcal{C}_n, \ldots, \mathcal{C}_{N-1})
\end{equation}
The routine $\textproc{ApplyMove}$ in \cref{algo:moves} performs exactly this action.

To access the gate and the index of a move, we use subscripts $G$ and $n$, respectively, e.g. if $M=(R^x_1, 4)$, then $M_G = R^x_1$ and $M_n = 4$. This definition allows us to conveniently pass around \emph{moves} between functions, which will be useful later.

For a circuit containing $N$ gates, all possible moves are given by
\begin{equation}
    \mathcal{M}_\mathrm{all} = \{(G, n)\:|\:0 \leq n \leq N~\mathrm{and}~G \in \mathcal{L}\}.
\end{equation}
However, many of the moves in $\mathcal{M}_\mathrm{all}$ will lead to redundancies in the circuit, because neighbouring identical gates acting on the same qubits can be merged straightforwardly. We eliminate modifications leading to such obvious redundancies and generate a set containing only potentially useful moves. The specific method we used is given in \cref{algo:moves} and works as follows. We separately look at each qubit $k$, and in one iteration only consider the gates in $\mathcal{C}$ acting on qubit $k$. Potentially useful modifications are then insertions of gates from the library between consecutive pairs of gates which also act on qubit $k$, but are different to the previous and following gate acting on qubit $k$.

In some cases -- e.g. if some gates in $\mathcal{L}$ commute with one another -- \cref{algo:moves} will still include moves that lead to redundancies. The quantum metric tensor as defined in \cref{sec:param_opt} can be used to detect such redundant gates. The details of this operation are given in the next subsection.

\subsection{Removal of superfluous gates} \label{sec:remove_gates}
In a circuit $\mathcal{C}$, not all gates in the sequence necessarily contribute to generating the desired unitary in a useful way. We use three distinct techniques with varying computational cost and ability to detect such redundancies, which we discuss in the following.

In our algorithms, we use these methods in the order \emph{small parameter}~$\rightarrow$~\emph{QMT-assisted} $\rightarrow$~\emph{trial and error}, as shown in \cref{algo:prune}. Each method can, in principle, also detect all redundancies of the previous methods\footnote{In the generalisation of also allowing non-parametrised gates in the circuit -- which we do not discuss further -- only \emph{trial and error} may be used for those specific gates.}, but is more costly to perform, which makes this staged approach useful.

\subsubsection{Small parameter removal}
The computationally cheapest way to detect non-contributing gates is by checking the associated parameters after optimising them.\footnote{We assume that every gate $G$ approaches the identity for small parameters, i.e. $\lim_{\theta\rightarrow 0}G(\theta) = \mathds{1}$.} For every circuit parameter close to 0 modulo\footnote{For practical reasons we use the slightly unusual definition of $a \bmod b = a - b\left\lfloor a/b + .5\right\rfloor$ which returns values in the interval $\left[-b/2, b/2\right)$ instead of the usual $[0, b)$.} $2\pi$, $\theta_k \approx 0 \bmod 2\pi$, the corresponding gate $\mathcal{C}_k$ can be removed immediately.

\subsubsection{Quantum metric tensor assisted removal}
As a more sophisticated and computationally slightly more expensive method to detect further redundancies, we check the quantum metric tensor (QMT) -- as defined in \cref{eq:QMT} -- for linearly dependent rows.

The QMT contains information about how the output state changes with respect to varying the parameters. If rows $i$ and $j$ in this tensor are linearly dependent, the linearised actions of $\mathcal{C}_i$ and $\mathcal{C}_j$ are equivalent in the tangent space of the circuit $\mathcal{C}$ at the current position $\vec{\theta}$ in parameter space. Intuitively, this means that changes to the parameters $\theta_i$ and $\theta_j$ from their current values would move the state in the same direction inside some subspace of the full Hilbert space. While this does not guarantee that $\mathcal{C}_i$ and $\mathcal{C}_j$ have equivalent actions in the full Hilbert space $\mathcal{H}$ or at a different point $\vec{\theta}'$ in parameter space, it is a strong indication of it. We therefore use 
\begin{equation} \label{eq:linearly_dependent}
    \big|(A_{i,\cdot}^\intercal \cdot A_{j,\cdot}) - \|A_{i,\cdot}\|\,\|A_{j,\cdot}\| \,\big| < \varepsilon_\mathrm{QMT},
\end{equation}
where $A_{i,\cdot}$ is the $i^\text{th}$ row vector of $\mat{A}$, as a heuristic for detecting potentially redundant gates with an appropriately small $\varepsilon_\mathrm{QMT}$.

At many points in the iteration the QMT is already known from the previously performed parameter optimisation and does not need to be explicitly re-calculated. As calculating this matrix is computationally relatively expensive, this saves valuable computing time.

The removal is performed as follows. For all pairs of rows $i,j$ we check the condition in \cref{eq:linearly_dependent}. If it is fulfilled, we adjust the parameter of the first gate to $\theta_i \gets \theta_i + \theta_j$ and remove the gate $\mathcal{C}_j$ from the circuit $\mathcal{C}$ as well as $\theta_j$ from the parameter vector $\vec{\theta}$. If this removal does not significantly increase the energy, the new circuit is kept, otherwise the deletion is reverted. Among other redundancies, this method allows the relatively easy detection of identical gates separated by gate sequences they (non-trivially) commute with.

\subsubsection{Trial and error removal}
The above method is good at finding redundancies where one gate can absorb a different one into its parameter. It cannot, however, detect cases where multiple gates need to adjust their parameters in order to compensate the removal of one specific gate. We therefore employ a third strategy, which is computationally more expensive, but can also detect much more subtle redundancies.

Given a circuit $\mathcal{C}$ and a set of gate indices $\mathcal{R}$ to be considered candidates for removal, we delete $\mathcal{C}_k$ for each $k\in\mathcal{R}$ from the circuit without replacement, and run a full parameter optimisation as in \cref{sec:param_opt} with the modified circuit. If the energy after the relaxation is not significantly higher than before, the deleted gate is considered redundant and remains removed. We refer to this method as a \emph{hard removal}, because the circuit is abruptly taken to a different point in parameter space, potentially far away from a local minimum.

An alternative to the aforementioned \emph{hard removal} explored by the authors but not reported here, is a method we refer to as \emph{soft removal}. The parameter $\theta_k$, for $k\in\mathcal{R}$, is shifted towards zero by some predetermined amount, and a single imaginary time step for all parameters except $\theta_k$ is performed right afterwards. This procedure is repeated until the parameter is close to zero, and only then is the gate completely removed. This allows the circuit to stay close to the local minimum it is already in. Therefore this method can occasionally lead to better results.

\subsection{Initial circuit} \label{sec:initial_circ}
When the gates in the considered library $\mathcal{L}$ have limited connectivity, i.e. not every qubit can interact with every other qubit directly, finding useful gate additions by randomly adding gates can become increasingly improbable. For example, in a linear chain of qubits with nearest-neighbour connectivity, having qubit 0 interact with qubit 3 requires the correct simultaneous addition of at least three gates. The effort of finding the correct combination of three gates is further hampered by the fact that -- at least for the artificial Hamiltonians $\tilde{H}$ we use for our circuit synthesis tasks -- the energy landscape with regard to adding only a subset of those gates is flat.

\begin{figure}[tbh]
\providecommand{\ket}[1]{\left|#1\right\rangle}
\begin{tikzpicture}[scale=1.000000,x=1pt,y=1pt]
\filldraw[color=white] (0.000000, -6.750000) rectangle (180.000000, 60.750000);
\draw[color=black] (0.000000,54.000000) -- (180.000000,54.000000);
\draw[color=black] (0.000000,40.500000) -- (180.000000,40.500000);
\draw[color=black] (0.000000,27.000000) -- (180.000000,27.000000);
\draw[color=black] (0.000000,13.500000) -- (180.000000,13.500000);
\draw[color=black] (0.000000,0.000000) -- (180.000000,0.000000);
\draw[densely dotted] (9.000000,54.000000) -- (9.000000,40.500000);
\begin{scope}
\draw (6.878680, 51.878680) -- (11.121320, 56.121320);
\draw (6.878680, 56.121320) -- (11.121320, 51.878680);
\end{scope}
\begin{scope}
\draw (6.878680, 38.378680) -- (11.121320, 42.621320);
\draw (6.878680, 42.621320) -- (11.121320, 38.378680);
\end{scope}
\draw[densely dotted] (45.000000,54.000000) -- (45.000000,40.500000);
\begin{scope}
\draw (42.878680, 51.878680) -- (47.121320, 56.121320);
\draw (42.878680, 56.121320) -- (47.121320, 51.878680);
\end{scope}
\begin{scope}
\draw (42.878680, 38.378680) -- (47.121320, 42.621320);
\draw (42.878680, 42.621320) -- (47.121320, 38.378680);
\end{scope}
\draw[densely dotted] (81.000000,54.000000) -- (81.000000,40.500000);
\begin{scope}
\draw (78.878680, 51.878680) -- (83.121320, 56.121320);
\draw (78.878680, 56.121320) -- (83.121320, 51.878680);
\end{scope}
\begin{scope}
\draw (78.878680, 38.378680) -- (83.121320, 42.621320);
\draw (78.878680, 42.621320) -- (83.121320, 38.378680);
\end{scope}
\draw[densely dotted] (117.000000,54.000000) -- (117.000000,40.500000);
\begin{scope}
\draw (114.878680, 51.878680) -- (119.121320, 56.121320);
\draw (114.878680, 56.121320) -- (119.121320, 51.878680);
\end{scope}
\begin{scope}
\draw (114.878680, 38.378680) -- (119.121320, 42.621320);
\draw (114.878680, 42.621320) -- (119.121320, 38.378680);
\end{scope}
\draw[densely dotted] (153.000000,54.000000) -- (153.000000,40.500000);
\begin{scope}
\draw (150.878680, 51.878680) -- (155.121320, 56.121320);
\draw (150.878680, 56.121320) -- (155.121320, 51.878680);
\end{scope}
\begin{scope}
\draw (150.878680, 38.378680) -- (155.121320, 42.621320);
\draw (150.878680, 42.621320) -- (155.121320, 38.378680);
\end{scope}
\draw[densely dotted] (9.000000,27.000000) -- (9.000000,13.500000);
\begin{scope}
\draw (6.878680, 24.878680) -- (11.121320, 29.121320);
\draw (6.878680, 29.121320) -- (11.121320, 24.878680);
\end{scope}
\begin{scope}
\draw (6.878680, 11.378680) -- (11.121320, 15.621320);
\draw (6.878680, 15.621320) -- (11.121320, 11.378680);
\end{scope}
\draw[densely dotted] (45.000000,27.000000) -- (45.000000,13.500000);
\begin{scope}
\draw (42.878680, 24.878680) -- (47.121320, 29.121320);
\draw (42.878680, 29.121320) -- (47.121320, 24.878680);
\end{scope}
\begin{scope}
\draw (42.878680, 11.378680) -- (47.121320, 15.621320);
\draw (42.878680, 15.621320) -- (47.121320, 11.378680);
\end{scope}
\draw[densely dotted] (81.000000,27.000000) -- (81.000000,13.500000);
\begin{scope}
\draw (78.878680, 24.878680) -- (83.121320, 29.121320);
\draw (78.878680, 29.121320) -- (83.121320, 24.878680);
\end{scope}
\begin{scope}
\draw (78.878680, 11.378680) -- (83.121320, 15.621320);
\draw (78.878680, 15.621320) -- (83.121320, 11.378680);
\end{scope}
\draw[densely dotted] (117.000000,27.000000) -- (117.000000,13.500000);
\begin{scope}
\draw (114.878680, 24.878680) -- (119.121320, 29.121320);
\draw (114.878680, 29.121320) -- (119.121320, 24.878680);
\end{scope}
\begin{scope}
\draw (114.878680, 11.378680) -- (119.121320, 15.621320);
\draw (114.878680, 15.621320) -- (119.121320, 11.378680);
\end{scope}
\draw[densely dotted] (153.000000,27.000000) -- (153.000000,13.500000);
\begin{scope}
\draw (150.878680, 24.878680) -- (155.121320, 29.121320);
\draw (150.878680, 29.121320) -- (155.121320, 24.878680);
\end{scope}
\begin{scope}
\draw (150.878680, 11.378680) -- (155.121320, 15.621320);
\draw (150.878680, 15.621320) -- (155.121320, 11.378680);
\end{scope}
\draw[densely dotted] (27.000000,40.500000) -- (27.000000,27.000000);
\begin{scope}
\draw (24.878680, 38.378680) -- (29.121320, 42.621320);
\draw (24.878680, 42.621320) -- (29.121320, 38.378680);
\end{scope}
\begin{scope}
\draw (24.878680, 24.878680) -- (29.121320, 29.121320);
\draw (24.878680, 29.121320) -- (29.121320, 24.878680);
\end{scope}
\draw[densely dotted] (63.000000,40.500000) -- (63.000000,27.000000);
\begin{scope}
\draw (60.878680, 38.378680) -- (65.121320, 42.621320);
\draw (60.878680, 42.621320) -- (65.121320, 38.378680);
\end{scope}
\begin{scope}
\draw (60.878680, 24.878680) -- (65.121320, 29.121320);
\draw (60.878680, 29.121320) -- (65.121320, 24.878680);
\end{scope}
\draw[densely dotted] (99.000000,40.500000) -- (99.000000,27.000000);
\begin{scope}
\draw (96.878680, 38.378680) -- (101.121320, 42.621320);
\draw (96.878680, 42.621320) -- (101.121320, 38.378680);
\end{scope}
\begin{scope}
\draw (96.878680, 24.878680) -- (101.121320, 29.121320);
\draw (96.878680, 29.121320) -- (101.121320, 24.878680);
\end{scope}
\draw[densely dotted] (135.000000,40.500000) -- (135.000000,27.000000);
\begin{scope}
\draw (132.878680, 38.378680) -- (137.121320, 42.621320);
\draw (132.878680, 42.621320) -- (137.121320, 38.378680);
\end{scope}
\begin{scope}
\draw (132.878680, 24.878680) -- (137.121320, 29.121320);
\draw (132.878680, 29.121320) -- (137.121320, 24.878680);
\end{scope}
\draw[densely dotted] (171.000000,40.500000) -- (171.000000,27.000000);
\begin{scope}
\draw (168.878680, 38.378680) -- (173.121320, 42.621320);
\draw (168.878680, 42.621320) -- (173.121320, 38.378680);
\end{scope}
\begin{scope}
\draw (168.878680, 24.878680) -- (173.121320, 29.121320);
\draw (168.878680, 29.121320) -- (173.121320, 24.878680);
\end{scope}
\draw[densely dotted] (27.000000,13.500000) -- (27.000000,0.000000);
\begin{scope}
\draw (24.878680, 11.378680) -- (29.121320, 15.621320);
\draw (24.878680, 15.621320) -- (29.121320, 11.378680);
\end{scope}
\begin{scope}
\draw (24.878680, -2.121320) -- (29.121320, 2.121320);
\draw (24.878680, 2.121320) -- (29.121320, -2.121320);
\end{scope}
\draw[densely dotted] (63.000000,13.500000) -- (63.000000,0.000000);
\begin{scope}
\draw (60.878680, 11.378680) -- (65.121320, 15.621320);
\draw (60.878680, 15.621320) -- (65.121320, 11.378680);
\end{scope}
\begin{scope}
\draw (60.878680, -2.121320) -- (65.121320, 2.121320);
\draw (60.878680, 2.121320) -- (65.121320, -2.121320);
\end{scope}
\draw[densely dotted] (99.000000,13.500000) -- (99.000000,0.000000);
\begin{scope}
\draw (96.878680, 11.378680) -- (101.121320, 15.621320);
\draw (96.878680, 15.621320) -- (101.121320, 11.378680);
\end{scope}
\begin{scope}
\draw (96.878680, -2.121320) -- (101.121320, 2.121320);
\draw (96.878680, 2.121320) -- (101.121320, -2.121320);
\end{scope}
\draw[densely dotted] (135.000000,13.500000) -- (135.000000,0.000000);
\begin{scope}
\draw (132.878680, 11.378680) -- (137.121320, 15.621320);
\draw (132.878680, 15.621320) -- (137.121320, 11.378680);
\end{scope}
\begin{scope}
\draw (132.878680, -2.121320) -- (137.121320, 2.121320);
\draw (132.878680, 2.121320) -- (137.121320, -2.121320);
\end{scope}
\draw[densely dotted] (171.000000,13.500000) -- (171.000000,0.000000);
\begin{scope}
\draw (168.878680, 11.378680) -- (173.121320, 15.621320);
\draw (168.878680, 15.621320) -- (173.121320, 11.378680);
\end{scope}
\begin{scope}
\draw (168.878680, -2.121320) -- (173.121320, 2.121320);
\draw (168.878680, 2.121320) -- (173.121320, -2.121320);
\end{scope}
\end{tikzpicture}
    \caption{The initial circuit we use for gate libraries $\mathcal{L}$ where two-qubit gates are limited to nearest neighbours.}
    \label{fig:substrate}
\end{figure}
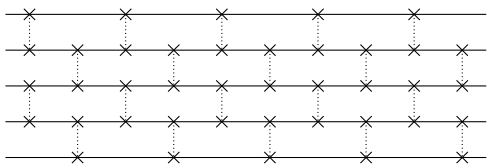

To alleviate these connectivity limitations, it can be helpful to choose not start with an empty circuit, but to have an initial structure of parameterised \textsc{swap} gates, where every qubit can be close to every other qubit at some point, and is able to -- but does not need to -- subsequently return to its original position. This can be achieved by an ansatz of the form
\begin{equation}
    \mathcal{C}^{(0)} = \prod_{n=0}^{N} \left[\prod_{k=1}^{\lceil N/2-1\rceil} E_{2k, 2k+1} \prod_{k=0}^{\lfloor N/2-1\rfloor} E_{2k, 2k+1} \right]
\end{equation}
where $N$ is the number of qubits and
\begin{equation} \label{eq:param_swap}
    E_{i, j} \coloneqq \exp(i\frac{\theta}{2} \textsc{swap}_{i, j})
\end{equation}
is a gate which performs no action for $\theta = 0$ and swaps the qubits $i$ and $j$ if $\theta = \pi$. The ansatz is visualised in \cref{fig:substrate}, where the $E$ gate is indicated by the usual \textsc{swap} symbol, but drawn with dotted lines.

The $E_{i,j}$ gates are typically not part of the gate library $\mathcal{L}$ and must be compiled to it separately. If all parameters converge to either $0$ or $\pi$, it is sufficient to compile the \textsc{swap} gate to the desired gate library $\mathcal{L}$.

\subsection{Circuit structure finding} \label{sec:algorithms}

Here we describe the algorithms we used to synthesise quantum circuits, all of which are adaptions of well-known optimisation techniques.

\subsubsection{Hill climbing}
The simplest and most straightforward algorithm we employ is a variant of hill climbing~\cite{skiena2020algorithm}, which iteratively searches some \emph{neighbourhood} of the current circuit structure, and accepts the lowest energy solution within that neighbourhood. In our case, we define the neighbourhood of a circuit structure as all circuits reachable by applying $N_\mathrm{moves}$ \emph{moves} as defined in \cref{sec:new_gates}. For hill climbing, we only consider $N_\mathrm{moves} = 1$.

Starting from some initial circuit $\mathcal{C}^{(0)}$, all moves are tried, their parameters optimised according to \cref{sec:param_opt}, and the energies of the resulting circuits are recorded. The move which resulted in the lowest energy is kept and becomes the new circuit $\mathcal{C}^{(1)}$. This procedure is repeated until the energy falls below a given threshold. Removal of non-contributing gates is only done once at the end of the iteration. \Cref{algo:hillclimb} shows this procedure.

Because it only checks the immediate neighbourhood of the current circuit for improvements, this method quickly gets stuck in local minima of the cost function. This problem is typical for hill climbing algorithms. In some very limited cases, it is possible to extend the search radius to all combinations of two or more sequential steps, i.e. $N_\mathrm{moves} > 1$, in order to escape such local optima. However, in most cases the search space volume grows very rapidly with the search depth, making larger search radii impractical.

\subsubsection{Random search}
To be able to escape local minima in which hill climbing gets stuck, we must make larger steps in configuration space. However, as mentioned in the previous section, the size of the neighbourhood grows too rapidly to exhaustively search it. We therefore resort to a variant of random search~\cite{skiena2020algorithm}, which in each iteration proposes a random modification to the circuit, and accepts it if it lowers the energy.

A single proposed modification -- we will call this a \emph{random step} -- consists of applying a number of $N_\mathrm{moves}$ randomly chosen \emph{moves} to the circuit, and only then optimising its parameters. This essentially means adding $N_\mathrm{moves}$ random gates to the circuit. Such a random step takes us further in the space of circuit structures and thus has the ability to escape from local optima. Because many of the added gates are potentially not contributing to the reduction in energy, we attempt to remove as many of the newly added gates as possible after each random step via the methods discussed in \cref{sec:remove_gates}.

When drawing a random move, some gate types (e.g.~controlled rotations) might be overrepresented compared to others (e.g.~local rotations), simply because there are more of them. This also skews the number of gates by type in the final circuit, which we found to sometimes hinder performance. We therefore first sort all moves into two groups, one containing only moves with single-qubit gates, and the other containing only those with two-qubit gates. To draw a random move, we first choose one of the groups with equal probability, and then uniformly draw a move from the chosen group. For clarity we omit this detail in \cref{algo:randomsearch}.

Instead of simply accepting every random step that lowers the cost function, we furthermore found it beneficial to sample a small number $N_\text{samp} \sim 10$ of random steps and choose to keep only the step resulting in the lowest cost. \Cref{algo:randomsearch} shows this procedure.

\subsubsection{Tabu search}
As an extension to random search -- which relies purely on chance to find useful modifications to the circuit -- we also used a variant of tabu search~\cite{glover1990tabu} to avoid repeated application of unsuccessful circuit modifications. Here we briefly outline the idea behind the algorithm and its potential pitfalls.

The overall structure of our variant of tabu search is exactly the same as for random search in \ref{algo:randomsearch}, but with $N_\mathrm{samp} = 1$. It has however, the additional feature of a \emph{tabu list}, which we call $\tau$. Whenever a move is performed, it is recorded into this list, together with a label recording at which iteration the move was performed. For a number of iterations $T_\text{tabu} \sim 20$, the same move must then not be repeated. Therefore, before choosing a random move, all tabu moves are removed from the set of potential moves.

To keep the action of the moves consistent when positions in the circuit change due to the insertion or deletion of gates, we update the insertion indices $M_n\:\:\forall\,M \in \tau$ in the stored moves accordingly, whenever the circuit changes. This means decrementing all stored indices $M_n > m\:\:\forall\,M \in \tau$ by 1 if the gate with index $m$ is deleted, and incrementing all indices $M_n \geq m\:\:\forall\,m \in \tau$ by 1 when a new gate at index $m$ is inserted.

\section{Results} \label{sec:results}

We present our numerical findings for different applications of our method in this section. Unless otherwise noted, the calculations use the default hyperparameters listed in \cref{tbl:hyperparams} in \cref{sec:hyperparams}.

\subsection{Cost functions as proxy for unitary equivalence} \label{sec:cost_results}

To determine whether two operators\footnote{From here on we will omit explicit indication of the parameter vector $\vec{\theta}$ wherever practical.} $\mathcal{C}$ and $U$ are equivalent, an appropriate metric to use is the global-phase invariant operator norm of their difference
\begin{align} \nonumber
    \mathcal{D}(U, \mathcal{C}) &= \min_\phi\,\|U - e^{i\phi}\mathcal{C}\| \\
    &= \min_\phi \left[\max_{\ket{\psi}} \|U\ket{\psi} - e^{i\phi}\mathcal{C}\ket{\psi}\|_2 \right], \label{eq:op_dist}
\end{align}
which we will refer to as the \emph{operator distance}. It is the maximum $L_2$-norm of the difference (and thus the Euclidean distance) between the desired output $U\ket{\psi}$ and the output of the recompiled version $\mathcal{C}\ket{\psi}$.

For the calculations only considering a subspace, we also define
\begin{equation}
    \mathcal{D}_\mathcal{S}(U, \mathcal{C}) \coloneqq \mathcal{D}(\Pi_\mathcal{S}\,U\,\Pi_\mathcal{S}, \Pi_\mathcal{S}\,\mathcal{C}\,\Pi_\mathcal{S})
\end{equation}
where $\Pi_\mathcal{S}$ is the projector onto the relevant subspace.

Throughout our results we use the expected energy of $H_\mathrm{sum}$ or $H_\mathrm{proj}$ as defined in \cref{eq:H_sum,eq:H_proj} to assess how closely a constructed circuit $\mathcal{C}$ reproduces the desired unitary $U$. Note, however, that in general $\expval*{\tilde{H}} \nsim \mathcal{D}(\mathcal{C}, U)$. This is easily demonstrated with an $n$-controlled Pauli-$Z$ gate $C_{1..n}[\sigma^z_{n+1}]$, for which the identity operator yields an energy of $\expval{H_\mathrm{proj}} = \expval{H_\mathrm{sum}} = 2^{-n}$, but the operator distance has its maximum value of $\mathcal{D}(\mathds{1}, C_{1..n}[\sigma^z_{n+1}])=2$.

\begin{figure}[tb]
\begin{tikzpicture}[trim axis right,baseline]

\definecolor{color0}{rgb}{0.12156862745098,0.466666666666667,0.705882352941177}
\definecolor{color1}{rgb}{1,0.317647058823529,0.282352941176471}
\definecolor{color2}{rgb}{1,0.498039215686275,0.0549019607843137}
\definecolor{color3}{rgb}{0.184313725490196,0.294117647058824,0.486274509803922}

\begin{axis}[height=.7\columnwidth, width=.96\columnwidth,
legend cell align={left},
legend style={fill opacity=0.8, draw opacity=1, text opacity=1, at={(0.03,0.97)}, anchor=north west, draw=white!80!black},
log basis x={10},
log basis y={10},
tick align=inside,
tick pos=left,
x grid style={white!69.0196078431373!black},
xlabel={$\expval*{\hat{H}}$},
xmin=4.98353679369512e-09, xmax=0.417909977296694,
xmode=log,
xtick style={color=black},
y grid style={white!69.0196078431373!black},
ylabel={$\mathcal{D}(U, \mathcal{C})$},
ymin=1.98369299868367e-18, ymax=0.101144683241379,
ymode=log,
ytick style={color=black}
]
\addplot [semithick, color0, mark=o, mark size=1, mark options={solid,draw=color1}, only marks]
table {%
3.14364946499168e-06 7.4e-12
0.08148309146539 0.0038
9.64666668934142e-05 5.86e-09
1.53560307533378e-05 1.48e-10
4.84028996134381e-06 1.33e-11
6.54159661109725e-06 1.28e-11
5.84376979323902e-06 2.86e-11
0.106031220382589 0.00413
1.30592500456845e-06 1.08e-12
1.50334726116046e-05 1.11e-10
0.0811835225521747 0.00386
3.28760912548668e-06 6.63e-12
0.149932009762781 0.0149
8.48655050826719e-05 4.34e-09
0.0985896227859147 0.00384
8.90560441237224e-05 4.91e-09
2.66184201075084e-06 4.19e-12
0.079694188006248 0.00385
0.00019535997220976 1.72e-08
0.000796714426248221 5.77e-07
0.0805510607550637 0.0038
3.14645619668764e-06 3.1e-12
2.3216530112714e-06 2.84e-12
7.3800867130017e-06 3.28e-11
0.000821868346246673 5.25e-07
0.000137545289010185 9.38e-09
9.64102116977927e-05 7.28e-09
0.0805513579360382 0.0038
0.0774557229546283 0.00381
1.84665684641298e-07 1.21e-14
0.000219368049317567 2.77e-08
2.9344911420866e-05 4.37e-10
1.93451415096313e-05 1.69e-10
8.52427135173324e-05 5.41e-09
2.00935539789432e-06 2.65e-12
8.27001120358563e-06 3.36e-11
0.0774032095105769 0.00382
1.14208600193705e-08 0
1.08384293417098e-06 6.94e-13
1.10844840190559e-06 4.36e-13
0.0396770968036764 0.000623
5.10932401947862e-06 1.45e-11
0.0528717678582107 0.00102
0.14347999895356 0.0084
2.25409792783628e-05 2.57e-10
0.000516221280270733 1.78e-07
0.152296286967427 0.0176
0.109194064643209 0.00477
9.0455519701815e-07 5.01e-13
0.000124972949730485 8.5e-09
4.59302536299286e-07 1.07e-13
0.178675938252537 0.011
0.182356647816191 0.012
6.43015014372536e-05 2.5e-09
4.18084237811444e-06 1.24e-11
5.9130177725278e-05 2.06e-09
2.58905615901351e-05 2.64e-10
3.69250842724517e-08 1.14e-17
8.81564956179301e-07 1.59e-13
4.56348253808339e-07 6.03e-14
0.00019300544081061 2.89e-08
8.17712683360185e-07 3.4e-13
6.93353457014129e-08 6.77e-17
6.8379942505914e-06 2.06e-11
0.0833796939335355 0.00362
4.45033766771537e-06 1.14e-11
5.91772598163629e-05 2.36e-09
7.96705837868829e-05 4.53e-09
0.0494079156704787 0.000963
0.049301930732634 0.000963
1.20263307641886e-07 1.89e-15
0.178938957837794 0.0111
5.93840654905695e-06 2.06e-11
1.41271657241076e-06 4.13e-13
6.55915743820254e-05 3.37e-09
9.47478851042207e-05 7.1e-09
8.71454941441214e-07 4e-13
8.69288933856277e-08 2.64e-15
3.55782805582508e-06 6.29e-12
0.10689959228442 0.00463
0.0803875212355764 0.0038
2.86636680560987e-05 4.1e-10
0.0239821864472292 0.000325
0.0492034015817418 0.000963
0.117660173785784 0.0048
0.0986542817721868 0.00385
0.084284201982103 0.00381
0.000312568220683049 7.89e-08
2.56675054473993e-07 1.77e-14
0.000189536390759682 2.18e-08
0.0750375078958324 0.00331
9.15709052442328e-05 7.03e-09
7.37888041379399e-06 2.72e-11
0.147645312316745 0.01
1.15759653627629e-06 5.94e-13
1.51724281665747e-06 1.24e-12
7.91551823239002e-05 3.59e-09
0.0799318150917624 0.0038
3.98152689205344e-05 8.91e-10
0.0859660444170338 0.00416
1.5089481712893e-06 8.17e-13
5.47749165496706e-05 3.09e-09
3.74596799178948e-06 5.46e-12
0.0991314942242094 0.00384
0.000375335863051218 9.12e-08
8.40447415666336e-05 3.83e-09
0.088409074324272 0.00416
1.05513661056954e-05 5.87e-11
9.12765895473762e-06 5e-11
2.74139687679744e-06 3.86e-12
0.0625065365933581 0.00223
0.181218711246723 0.0156
4.10075275147048e-07 7.93e-14
4.88768399733708e-06 1.04e-11
6.69098796805405e-08 3.67e-16
1.40771131397354e-06 1.05e-12
0.179192722072697 0.0114
1.45961270174152e-06 1.26e-12
0.121270879843139 0.00413
4.09046713763622e-06 9.67e-12
};
\addlegendentry{$H_\mathrm{sum}$}
\addplot [semithick, color2, mark=o, mark size=1, mark options={solid,draw=color3}, only marks]
table {%
3.14364946499168e-06 5.80013814754921e-12
0.08148309146539 0.00295400417565539
9.64666668934142e-05 4.28010804576218e-09
1.53560307533378e-05 9.28258581112118e-11
4.84028996134381e-06 1.02303721050134e-11
6.54159661109725e-06 1.76576531174533e-11
5.84376979323902e-06 1.79918302478654e-11
0.106031220382589 0.00498828130358142
1.30592500456845e-06 8.10462807976364e-13
1.50334726116046e-05 8.77978800772894e-11
0.0811835225521747 0.00300009596427819
3.28760912548668e-06 4.56445992114141e-12
0.149932009762781 0.0111744898019234
8.48655050826719e-05 3.45294892678538e-09
0.0985896227859147 0.00481537532131104
8.90560441237224e-05 3.7361913562961e-09
2.66184201075084e-06 3.02635694282571e-12
0.079694188006248 0.00297373543727453
0.00019535997220976 1.42979629247364e-08
0.000796714426248221 3.17417004569442e-07
0.0805510607550637 0.00295223548563683
3.14645619668764e-06 3.65019126036259e-12
2.3216530112714e-06 2.25430785150138e-12
7.3800867130017e-06 2.1637691638432e-11
0.000821868346246673 3.40590226022286e-07
0.000137545289010185 6.78420553068548e-09
9.64102116977927e-05 5.11206965647659e-09
0.0805513579360382 0.0029521921692558
0.0774557229546283 0.00294739300518798
1.84665684641298e-07 1.62092561595273e-14
0.000219368049317567 1.83569095479896e-08
2.9344911420866e-05 3.54081763909164e-10
1.93451415096313e-05 1.41830991395864e-10
8.52427135173324e-05 4.11229317265338e-09
2.00935539789432e-06 1.81643589058922e-12
8.27001120358563e-06 2.71631606096889e-11
0.0774032095105769 0.0029468960175566
1.14208600193705e-08 1.22124532708767e-15
1.08384293417098e-06 5.55444579219966e-13
1.10844840190559e-06 3.1818991885757e-13
0.0396770968036764 0.000608442097838058
5.10932401947862e-06 1.16776588399148e-11
0.0528717678582107 0.00124265207296614
0.14347999895356 0.00793736171150849
2.25409792783628e-05 2.27765251104017e-10
0.000516221280270733 1.2816665440063e-07
0.152296286967427 0.0144115193527523
0.109194064643209 0.00415064961594924
9.0455519701815e-07 3.93463039927155e-13
0.000124972949730485 5.28554455758723e-09
4.59302536299286e-07 9.12603326241879e-14
0.178675938252537 0.0117357097106957
0.182356647816191 0.0130978829606897
6.43015014372536e-05 2.01368755003983e-09
4.18084237811444e-06 9.32720567448087e-12
5.9130177725278e-05 1.48854373271945e-09
2.58905615901351e-05 3.27599281035873e-10
3.69250842724517e-08 1.22124532708767e-15
8.81564956179301e-07 3.3717473257866e-13
4.56348253808339e-07 1.00808250635964e-13
0.00019300544081061 1.81923262010386e-08
8.17712683360185e-07 2.75224287804576e-13
6.93353457014129e-08 1.11022302462516e-15
6.8379942505914e-06 2.01809680078213e-11
0.0833796939335355 0.00396439313877162
4.45033766771537e-06 9.2512664195965e-12
5.91772598163629e-05 1.50973922252717e-09
7.96705837868829e-05 3.26493687641261e-09
0.0494079156704787 0.0012047255687081
0.049301930732634 0.00120475022653121
1.20263307641886e-07 4.10782519111308e-15
0.178938957837794 0.0119790762866102
5.93840654905695e-06 1.59546820199807e-11
1.41271657241076e-06 9.47575351517571e-13
6.55915743820254e-05 2.07897687953817e-09
9.47478851042207e-05 4.61917948335611e-09
8.71454941441214e-07 3.42392780794398e-13
8.69288933856277e-08 4.88498130835069e-15
3.55782805582508e-06 5.26800825184637e-12
0.10689959228442 0.00387795803528845
0.0803875212355764 0.00295445132753025
2.86636680560987e-05 2.90148127746193e-10
0.0239821864472292 0.000210559160670143
0.0492034015817418 0.00120456614844922
0.117660173785784 0.00601486244384941
0.0986542817721868 0.00481861185940191
0.084284201982103 0.0029589806819651
0.000312568220683049 4.76892090262027e-08
2.56675054473993e-07 2.36477504245158e-14
0.000189536390759682 1.45199808843088e-08
0.0750375078958324 0.00248373634828902
9.15709052442328e-05 7.16529358069096e-09
7.37888041379399e-06 2.38964403820319e-11
0.147645312316745 0.00900772470525935
1.15759653627629e-06 7.56616991282044e-13
1.51724281665747e-06 1.01652020134679e-12
7.91551823239002e-05 2.8843119004307e-09
0.0799318150917624 0.00294680998366603
3.98152689205344e-05 6.28973095828655e-10
0.0859660444170338 0.0032396542513351
1.5089481712893e-06 9.68114477473137e-13
5.47749165496706e-05 1.76636427706711e-09
3.74596799178948e-06 5.80102632596891e-12
0.0991314942242094 0.00481628134615986
0.000375335863051218 5.85795667529254e-08
8.40447415666336e-05 3.42089567784143e-09
0.088409074324272 0.00323409193648605
1.05513661056954e-05 4.98945329496792e-11
9.12765895473762e-06 4.52740067657942e-11
2.74139687679744e-06 3.3566482926517e-12
0.0625065365933581 0.00166466492612105
0.181218711246723 0.0107619904776263
4.10075275147048e-07 6.92779167366098e-14
4.88768399733708e-06 7.8607120812535e-12
6.69098796805405e-08 1.22124532708767e-15
1.40771131397354e-06 8.65973959207622e-13
0.179192722072697 0.0125614220126092
1.45961270174152e-06 1.0902390101819e-12
0.121270879843139 0.00539172813760591
4.09046713763622e-06 6.36501962247848e-12
};
\addlegendentry{$H_\mathrm{proj}$}
\end{axis}

\pgfresetboundingbox
\useasboundingbox
(-1.7,-1.2) rectangle (6.8,4.6);


\end{tikzpicture}
    \caption{Operator distance between the target $U$ and recompiled $\mathcal{C}$ unitaries for a set of recompiled QFT circuits on 5 qubits. The data points at very small costs are in the range of numerical noise and likely influenced by finite data type accuracy.}
    \label{fig:op_dist_cost}
\end{figure}
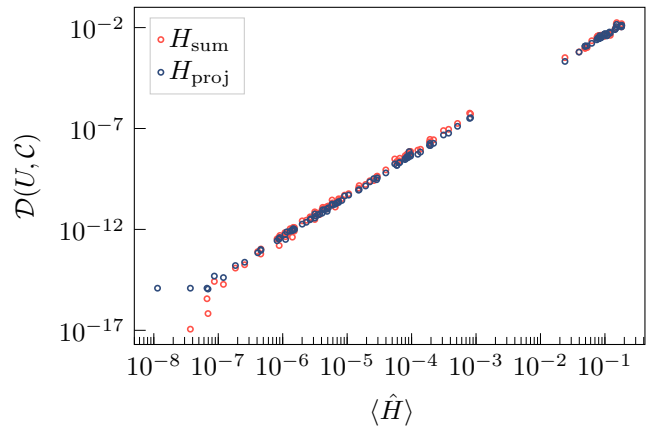

We therefore surveyed the resulting circuits of many 5-qubit QFT synthesis results and compared the cost functions of $\expval{H_\mathrm{sum}}$ and $\expval{H_\mathrm{proj}}$ to the actual operator distance $\mathcal{D}$. The results are plotted in \cref{fig:op_dist_cost} and show that close to convergence the used cost functions are both very good proxies for the actual operator distance. We therefore confidently use our cost functions to assess the quality of the recompiled circuit.

\subsection{Random unitaries}

\subsubsection{Dense unitaries}
\begin{figure}
\begin{tikzpicture}

\definecolor{darkgray176}{RGB}{176,176,176}
\definecolor{lightgray204}{RGB}{204,204,204}
\definecolor{steelblue39125161}{RGB}{39,125,161}
\definecolor{tomato2496568}{RGB}{249,65,68}

\begin{axis}[
axis on top,
grid style={line width=.5pt},
height=14em,
xtick={3,4},
legend cell align={left},
legend style={
  fill opacity=0.8,
  draw opacity=1,
  text opacity=1,
  at={(0.03,0.97)},
  anchor=north west,
  draw=lightgray204
},
tick align=outside,
tick pos=left,
unbounded coords=jump,
width=\columnwidth,
x grid style={darkgray176},
xlabel={\(\displaystyle n_\mathrm{qb}\)},
x label style={yshift=1.2em},
xmajorgrids,
xmin=2.29, xmax=4.71,
xtick style={color=black},
y grid style={darkgray176},
ylabel={$n_\mathrm{gates}$},
y label style={yshift=-2.1em},
ymin=0, ymax=320,
ytick={0,60,260,320},
ytick style={color=black}
]
\path [fill=tomato2496568, fill opacity=0.6]
(axis cs:3,61.5)
--(axis cs:3,61.5)
--(axis cs:3,61.5)
--(axis cs:3,61.5)
--cycle;
\path [fill=tomato2496568, fill opacity=0.6]
(axis cs:3,61.5)
--(axis cs:3,61.5)
--(axis cs:2.75820895522388,61.5)
--(axis cs:2.75820895522388,64.5)
--(axis cs:2.4,64.5)
--(axis cs:2.4,67.5)
--(axis cs:2.94626865671642,67.5)
--(axis cs:2.94626865671642,70.5)
--(axis cs:3,70.5)
--(axis cs:3,70.5)
--(axis cs:3,70.5)
--(axis cs:3,70.5)
--(axis cs:3,67.5)
--(axis cs:3,67.5)
--(axis cs:3,64.5)
--(axis cs:3,64.5)
--(axis cs:3,61.5)
--(axis cs:3,61.5)
--cycle;
\path [fill=tomato2496568, fill opacity=0.6]
(axis cs:3,73.5)
--(axis cs:3,73.5)
--(axis cs:3,73.5)
--(axis cs:3,73.5)
--cycle;
\path [fill=tomato2496568, fill opacity=0.6]
(axis cs:3,73.5)
--(axis cs:3,73.5)
--(axis cs:3,73.5)
--(axis cs:3,73.5)
--cycle;

\path [fill=steelblue39125161, fill opacity=0.6]
(axis cs:3,61.5)
--(axis cs:3,61.5)
--(axis cs:3,61.5)
--(axis cs:3,61.5)
--cycle;
\path [fill=steelblue39125161, fill opacity=0.6]
(axis cs:3,61.5)
--(axis cs:3,61.5)
--(axis cs:3.33134328358209,61.5)
--(axis cs:3.33134328358209,64.5)
--(axis cs:3.54626865671642,64.5)
--(axis cs:3.54626865671642,67.5)
--(axis cs:3.01791044776119,67.5)
--(axis cs:3.01791044776119,70.5)
--(axis cs:3,70.5)
--(axis cs:3,70.5)
--(axis cs:3,70.5)
--(axis cs:3,70.5)
--(axis cs:3,67.5)
--(axis cs:3,67.5)
--(axis cs:3,64.5)
--(axis cs:3,64.5)
--(axis cs:3,61.5)
--(axis cs:3,61.5)
--cycle;
\path [fill=steelblue39125161, fill opacity=0.6]
(axis cs:3,73.5)
--(axis cs:3,73.5)
--(axis cs:3,73.5)
--(axis cs:3,73.5)
--cycle;
\path [fill=steelblue39125161, fill opacity=0.6]
(axis cs:3,73.5)
--(axis cs:3,73.5)
--(axis cs:3,73.5)
--(axis cs:3,73.5)
--cycle;

\path [fill=tomato2496568, fill opacity=0.6]
(axis cs:4,61.5)
--(axis cs:4,61.5)
--(axis cs:4,61.5)
--(axis cs:4,61.5)
--cycle;
\path [fill=tomato2496568, fill opacity=0.6]
(axis cs:4,61.5)
--(axis cs:4,61.5)
--(axis cs:4,61.5)
--(axis cs:4,61.5)
--cycle;
\path [fill=tomato2496568, fill opacity=0.6]
(axis cs:4,61.5)
--(axis cs:4,61.5)
--(axis cs:4,61.5)
--(axis cs:4,61.5)
--cycle;
\path [fill=tomato2496568, fill opacity=0.6]
(axis cs:4,262.5)
--(axis cs:4,262.5)
--(axis cs:3.98235294117647,262.5)
--(axis cs:3.98235294117647,265.5)
--(axis cs:3.94705882352941,265.5)
--(axis cs:3.94705882352941,268.5)
--(axis cs:3.82352941176471,268.5)
--(axis cs:3.82352941176471,271.5)
--(axis cs:3.61176470588235,271.5)
--(axis cs:3.61176470588235,274.5)
--(axis cs:3.48823529411765,274.5)
--(axis cs:3.48823529411765,277.5)
--(axis cs:3.62941176470588,277.5)
--(axis cs:3.62941176470588,280.5)
--(axis cs:3.75294117647059,280.5)
--(axis cs:3.75294117647059,283.5)
--(axis cs:4,283.5)
--(axis cs:4,283.5)
--(axis cs:4,283.5)
--(axis cs:4,283.5)
--(axis cs:4,280.5)
--(axis cs:4,280.5)
--(axis cs:4,277.5)
--(axis cs:4,277.5)
--(axis cs:4,274.5)
--(axis cs:4,274.5)
--(axis cs:4,271.5)
--(axis cs:4,271.5)
--(axis cs:4,268.5)
--(axis cs:4,268.5)
--(axis cs:4,265.5)
--(axis cs:4,265.5)
--(axis cs:4,262.5)
--(axis cs:4,262.5)
--cycle;
\path [fill=tomato2496568, fill opacity=0.6]
(axis cs:4,286.5)
--(axis cs:4,286.5)
--(axis cs:4,286.5)
--(axis cs:4,286.5)
--cycle;
\path [fill=tomato2496568, fill opacity=0.6]
(axis cs:4,286.5)
--(axis cs:4,286.5)
--(axis cs:4,286.5)
--(axis cs:4,286.5)
--cycle;

\path [fill=steelblue39125161, fill opacity=0.6]
(axis cs:4,61.5)
--(axis cs:4,61.5)
--(axis cs:4,61.5)
--(axis cs:4,61.5)
--cycle;
\path [fill=steelblue39125161, fill opacity=0.6]
(axis cs:4,61.5)
--(axis cs:4,61.5)
--(axis cs:4,61.5)
--(axis cs:4,61.5)
--cycle;
\path [fill=steelblue39125161, fill opacity=0.6]
(axis cs:4,61.5)
--(axis cs:4,61.5)
--(axis cs:4,61.5)
--(axis cs:4,61.5)
--cycle;
\path [fill=steelblue39125161, fill opacity=0.6]
(axis cs:4,265.5)
--(axis cs:4,265.5)
--(axis cs:4.03529411764706,265.5)
--(axis cs:4.03529411764706,268.5)
--(axis cs:4.22941176470588,268.5)
--(axis cs:4.22941176470588,271.5)
--(axis cs:4.45882352941176,271.5)
--(axis cs:4.45882352941176,274.5)
--(axis cs:4.6,274.5)
--(axis cs:4.6,277.5)
--(axis cs:4.24705882352941,277.5)
--(axis cs:4.24705882352941,280.5)
--(axis cs:4.14117647058824,280.5)
--(axis cs:4.14117647058824,283.5)
--(axis cs:4.03529411764706,283.5)
--(axis cs:4.03529411764706,286.5)
--(axis cs:4.01764705882353,286.5)
--(axis cs:4.01764705882353,289.5)
--(axis cs:4,289.5)
--(axis cs:4,289.5)
--(axis cs:4,289.5)
--(axis cs:4,289.5)
--(axis cs:4,286.5)
--(axis cs:4,286.5)
--(axis cs:4,283.5)
--(axis cs:4,283.5)
--(axis cs:4,280.5)
--(axis cs:4,280.5)
--(axis cs:4,277.5)
--(axis cs:4,277.5)
--(axis cs:4,274.5)
--(axis cs:4,274.5)
--(axis cs:4,271.5)
--(axis cs:4,271.5)
--(axis cs:4,268.5)
--(axis cs:4,268.5)
--(axis cs:4,265.5)
--(axis cs:4,265.5)
--cycle;

\addplot [line width=0.5pt, tomato2496568, forget plot]
table {%
3 61.5
nan nan
3 61.5
2.75820895522388 61.5
2.75820895522388 64.5
2.4 64.5
2.4 67.5
2.94626865671642 67.5
2.94626865671642 70.5
3 70.5
nan nan
3 73.5
nan nan
3 73.5
};
\addplot [line width=0.5pt, steelblue39125161, forget plot]
table {%
3 61.5
nan nan
3 61.5
3.33134328358209 61.5
3.33134328358209 64.5
3.54626865671642 64.5
3.54626865671642 67.5
3.01791044776119 67.5
3.01791044776119 70.5
3 70.5
nan nan
3 73.5
nan nan
3 73.5
};
\addplot [line width=0.5pt, tomato2496568, forget plot]
table {%
4 61.5
nan nan
4 61.5
nan nan
4 61.5
nan nan
4 262.5
3.98235294117647 262.5
3.98235294117647 265.5
3.94705882352941 265.5
3.94705882352941 268.5
3.82352941176471 268.5
3.82352941176471 271.5
3.61176470588235 271.5
3.61176470588235 274.5
3.48823529411765 274.5
3.48823529411765 277.5
3.62941176470588 277.5
3.62941176470588 280.5
3.75294117647059 280.5
3.75294117647059 283.5
4 283.5
nan nan
4 286.5
nan nan
4 286.5
};
\addplot [line width=0.5pt, steelblue39125161, forget plot]
table {%
4 61.5
nan nan
4 61.5
nan nan
4 61.5
nan nan
4 265.5
4.03529411764706 265.5
4.03529411764706 268.5
4.22941176470588 268.5
4.22941176470588 271.5
4.45882352941176 271.5
4.45882352941176 274.5
4.6 274.5
4.6 277.5
4.24705882352941 277.5
4.24705882352941 280.5
4.14117647058824 280.5
4.14117647058824 283.5
4.03529411764706 283.5
4.03529411764706 286.5
4.01764705882353 286.5
4.01764705882353 289.5
4 289.5
};
\end{axis}

\pgfresetboundingbox
\useasboundingbox
(-.85,-.7) rectangle (7.15,3.5);


\end{tikzpicture}
    \caption{Histograms of the number of gates required to express Haar-distributed unitaries as a circuit containing only gates from $\mathcal{L}_\mathrm{allrot}$ for different numbers of qubits. Left (red) data was obtained using tabu search, right (blue) data is from our random search algorithm.}
    \label{fig:random_u_results}
\end{figure}
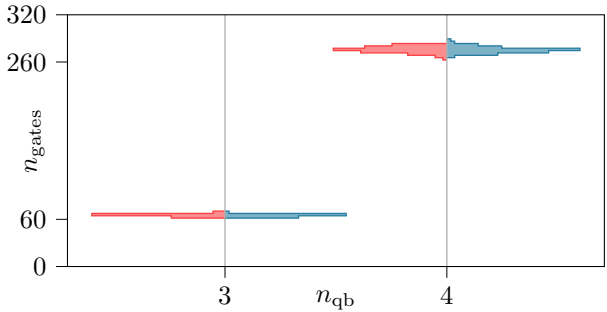

To test and benchmark our circuit synthesis protocols, we generated uniformly Haar-distributed random unitaries\footnote{The unitaries were created using \texttt{scipy.stats.unitary\_group} in Python and \texttt{CircularUnitaryMatrixDistribution} in Mathematica.} as targets, and tried to find circuit representations for them. For each target on $N$ qubits we ran a single attempt to create a circuit performing the same action using the gate set of
\begin{multline}
    \mathcal{L}_\mathrm{allrot} = \{R^\sigma_k, C_\ell[R^\sigma_k] \:|\: \sigma \in \{x, y, z\}, \\ k, \ell \in [N] \text{ and } k \neq \ell\}
\end{multline}
where $[N] \equiv \{1,\ldots,N\}$, which contains all local single-qubit Pauli rotations $R^x_k$, $R^y_k$, and $R^z_k$ on each qubit $k$, as well as single-controlled versions thereof with no connectivity constraints. The results of 100 random unitaries per method and number of qubits are shown in \cref{fig:random_u_results}.

We find that the number of gates in the circuits produced by random and tabu search correlates well with the increasing degrees of freedom of the targets, which for a dense unitary on $N$ qubits is $2^{2N}$. Therefore, for 3 qubits, we would on average expect to require no fewer than 64 gates, and find a mean number of 65 gates in our synthesised circuits. For 4 qubits, we observe a mean of 275 gates, where at least 256 would be expected. This indicates that, at least for dense, unstructured unitaries, the circuits produced by tabu and random search do not contain a large number of superfluous gates.

Note that the results show no significant difference between tabu and random search, which we discuss in \cref{sec:qft_results}.

\subsubsection{Subspace compilation}

\begin{figure*}[tb]
    \input{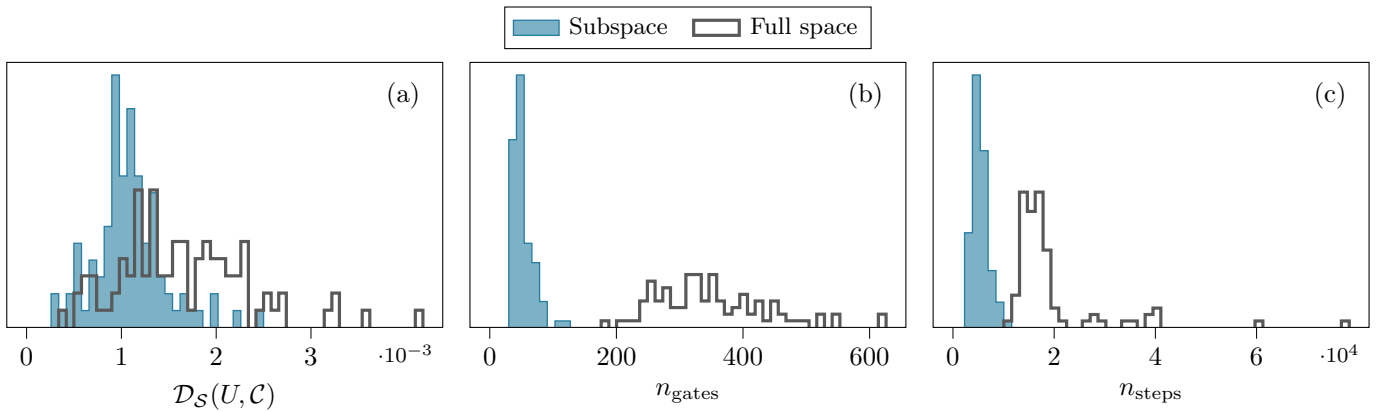}
    \vspace{-3ex}
    \caption{Normalised histograms showing properties of synthesised circuits $\mathcal{C}$ recovering the action of single 4-state block $U_1$ within a block-diagonal operator $U$ on 4 qubits. The blocks making up $U$ are each Haar distributed random unitaries. Empty histograms in grey \ref{leg:fullspace} show results for the full space method, filled histograms in blue \ref{leg:subspace} show subspace results. Circuits for same unitary $U$ were synthesised 100 times. (a) Operator distance as in \cref{eq:op_dist} within the subspace $\mathcal{H}_1$. (b) Number of gates in the resulting circuit. (c) Number of total imaginary time evolution steps needed by the algorithm.}
    \label{fig:subspace}
\end{figure*}

In order to assess the efficacy of compiling a unitary only in a certain subspace, we generated a random unitary operator which is block-diagonal in the computational basis when sorted by Hamming weight, i.e. the number of ones in each state belonging to the same block is equal. Each of these blocks was then assigned a Haar-distributed random unitary. We synthesised circuits representing the full unitary, as well as only the block with a Hamming weight of 1, i.e. the states $\{\ket{0001}, \ket{0010}, \ket{0100}, \ket{1000}\}$, using random search. For numerical reasons -- and because at this problem size barren plateaus proved not to be an issue -- we employed $\expval*{H_\mathrm{proj}}$ as our cost function in this example. The target energy was set to $\expval*{H_\mathrm{proj}}\leq 10^{-5}$, and other hyperparameters were $N_\mathrm{moves} = 30$ and $N_\mathrm{samp} = 10$. Because of the stochastic nature the compilation process, we synthesised 100 circuits each in the full- and subspace, of which 97 of the subspace and 95 of the full space attempts converged. \Cref{fig:subspace} summarises the outcomes of the successful calculations.

The results distinctly show that for comparable accuracy (a), considering only a subspace results in vastly fewer gates (b), as well as many fewer iterations (c). Therefore we expect this method to yield much better results whenever the target unitary conserves some quantity, and the relevant subspace is known in advance. Prominently, this is the case for time evolution operators in quantum chemistry, where the number of electrons is a conserved and known quantity, and corresponds to the Hamming weight of the states. The sister paper to the present work~\cite{gustiani2022variational} explores this application in greater detail.

\subsection{Quantum Fourier transform} \label{sec:qft_results}
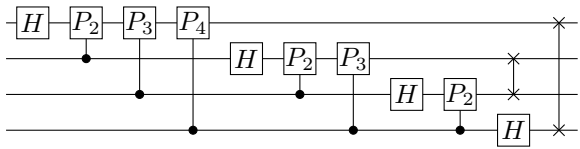
\begin{figure}[b]
\providecommand{\ket}[1]{\left|#1\right\rangle}
\begin{tikzpicture}[scale=1.000000,x=1pt,y=1pt]
\filldraw[color=white] (0.000000, -6.750000) rectangle (214.000000, 47.250000);
\draw[color=black] (0.000000,40.500000) -- (214.000000,40.500000);
\draw[color=black] (0.000000,27.000000) -- (214.000000,27.000000);
\draw[color=black] (0.000000,13.500000) -- (214.000000,13.500000);
\draw[color=black] (0.000000,0.000000) -- (214.000000,0.000000);
\begin{scope}
\draw[fill=white] (10.000000, 40.500000) +(-45.000000:8.485281pt and 8.485281pt) -- +(45.000000:8.485281pt and 8.485281pt) -- +(135.000000:8.485281pt and 8.485281pt) -- +(225.000000:8.485281pt and 8.485281pt) -- cycle;
\clip (10.000000, 40.500000) +(-45.000000:8.485281pt and 8.485281pt) -- +(45.000000:8.485281pt and 8.485281pt) -- +(135.000000:8.485281pt and 8.485281pt) -- +(225.000000:8.485281pt and 8.485281pt) -- cycle;
\draw (10.000000, 40.500000) node {$H$};
\end{scope}
\draw (30.000000,40.500000) -- (30.000000,27.000000);
\begin{scope}
\draw[fill=white] (30.000000, 40.500000) +(-45.000000:8.485281pt and 8.485281pt) -- +(45.000000:8.485281pt and 8.485281pt) -- +(135.000000:8.485281pt and 8.485281pt) -- +(225.000000:8.485281pt and 8.485281pt) -- cycle;
\clip (30.000000, 40.500000) +(-45.000000:8.485281pt and 8.485281pt) -- +(45.000000:8.485281pt and 8.485281pt) -- +(135.000000:8.485281pt and 8.485281pt) -- +(225.000000:8.485281pt and 8.485281pt) -- cycle;
\draw (30.000000, 40.500000) node {$P_2$};
\end{scope}
\filldraw (30.000000, 27.000000) circle(1.500000pt);
\draw (50.000000,40.500000) -- (50.000000,13.500000);
\begin{scope}
\draw[fill=white] (50.000000, 40.500000) +(-45.000000:8.485281pt and 8.485281pt) -- +(45.000000:8.485281pt and 8.485281pt) -- +(135.000000:8.485281pt and 8.485281pt) -- +(225.000000:8.485281pt and 8.485281pt) -- cycle;
\clip (50.000000, 40.500000) +(-45.000000:8.485281pt and 8.485281pt) -- +(45.000000:8.485281pt and 8.485281pt) -- +(135.000000:8.485281pt and 8.485281pt) -- +(225.000000:8.485281pt and 8.485281pt) -- cycle;
\draw (50.000000, 40.500000) node {$P_3$};
\end{scope}
\filldraw (50.000000, 13.500000) circle(1.500000pt);
\draw (70.000000,40.500000) -- (70.000000,0.000000);
\begin{scope}
\draw[fill=white] (70.000000, 40.500000) +(-45.000000:8.485281pt and 8.485281pt) -- +(45.000000:8.485281pt and 8.485281pt) -- +(135.000000:8.485281pt and 8.485281pt) -- +(225.000000:8.485281pt and 8.485281pt) -- cycle;
\clip (70.000000, 40.500000) +(-45.000000:8.485281pt and 8.485281pt) -- +(45.000000:8.485281pt and 8.485281pt) -- +(135.000000:8.485281pt and 8.485281pt) -- +(225.000000:8.485281pt and 8.485281pt) -- cycle;
\draw (70.000000, 40.500000) node {$P_4$};
\end{scope}
\filldraw (70.000000, 0.000000) circle(1.500000pt);
\begin{scope}
\draw[fill=white] (90.000000, 27.000000) +(-45.000000:8.485281pt and 8.485281pt) -- +(45.000000:8.485281pt and 8.485281pt) -- +(135.000000:8.485281pt and 8.485281pt) -- +(225.000000:8.485281pt and 8.485281pt) -- cycle;
\clip (90.000000, 27.000000) +(-45.000000:8.485281pt and 8.485281pt) -- +(45.000000:8.485281pt and 8.485281pt) -- +(135.000000:8.485281pt and 8.485281pt) -- +(225.000000:8.485281pt and 8.485281pt) -- cycle;
\draw (90.000000, 27.000000) node {$H$};
\end{scope}
\draw (110.000000,27.000000) -- (110.000000,13.500000);
\begin{scope}
\draw[fill=white] (110.000000, 27.000000) +(-45.000000:8.485281pt and 8.485281pt) -- +(45.000000:8.485281pt and 8.485281pt) -- +(135.000000:8.485281pt and 8.485281pt) -- +(225.000000:8.485281pt and 8.485281pt) -- cycle;
\clip (110.000000, 27.000000) +(-45.000000:8.485281pt and 8.485281pt) -- +(45.000000:8.485281pt and 8.485281pt) -- +(135.000000:8.485281pt and 8.485281pt) -- +(225.000000:8.485281pt and 8.485281pt) -- cycle;
\draw (110.000000, 27.000000) node {$P_2$};
\end{scope}
\filldraw (110.000000, 13.500000) circle(1.500000pt);
\draw (130.000000,27.000000) -- (130.000000,0.000000);
\begin{scope}
\draw[fill=white] (130.000000, 27.000000) +(-45.000000:8.485281pt and 8.485281pt) -- +(45.000000:8.485281pt and 8.485281pt) -- +(135.000000:8.485281pt and 8.485281pt) -- +(225.000000:8.485281pt and 8.485281pt) -- cycle;
\clip (130.000000, 27.000000) +(-45.000000:8.485281pt and 8.485281pt) -- +(45.000000:8.485281pt and 8.485281pt) -- +(135.000000:8.485281pt and 8.485281pt) -- +(225.000000:8.485281pt and 8.485281pt) -- cycle;
\draw (130.000000, 27.000000) node {$P_3$};
\end{scope}
\filldraw (130.000000, 0.000000) circle(1.500000pt);
\begin{scope}
\draw[fill=white] (150.000000, 13.500000) +(-45.000000:8.485281pt and 8.485281pt) -- +(45.000000:8.485281pt and 8.485281pt) -- +(135.000000:8.485281pt and 8.485281pt) -- +(225.000000:8.485281pt and 8.485281pt) -- cycle;
\clip (150.000000, 13.500000) +(-45.000000:8.485281pt and 8.485281pt) -- +(45.000000:8.485281pt and 8.485281pt) -- +(135.000000:8.485281pt and 8.485281pt) -- +(225.000000:8.485281pt and 8.485281pt) -- cycle;
\draw (150.000000, 13.500000) node {$H$};
\end{scope}
\draw (170.000000,13.500000) -- (170.000000,0.000000);
\begin{scope}
\draw[fill=white] (170.000000, 13.500000) +(-45.000000:8.485281pt and 8.485281pt) -- +(45.000000:8.485281pt and 8.485281pt) -- +(135.000000:8.485281pt and 8.485281pt) -- +(225.000000:8.485281pt and 8.485281pt) -- cycle;
\clip (170.000000, 13.500000) +(-45.000000:8.485281pt and 8.485281pt) -- +(45.000000:8.485281pt and 8.485281pt) -- +(135.000000:8.485281pt and 8.485281pt) -- +(225.000000:8.485281pt and 8.485281pt) -- cycle;
\draw (170.000000, 13.500000) node {$P_2$};
\end{scope}
\filldraw (170.000000, 0.000000) circle(1.500000pt);
\begin{scope}
\draw[fill=white] (190.000000, -0.000000) +(-45.000000:8.485281pt and 8.485281pt) -- +(45.000000:8.485281pt and 8.485281pt) -- +(135.000000:8.485281pt and 8.485281pt) -- +(225.000000:8.485281pt and 8.485281pt) -- cycle;
\clip (190.000000, -0.000000) +(-45.000000:8.485281pt and 8.485281pt) -- +(45.000000:8.485281pt and 8.485281pt) -- +(135.000000:8.485281pt and 8.485281pt) -- +(225.000000:8.485281pt and 8.485281pt) -- cycle;
\draw (190.000000, -0.000000) node {$H$};
\end{scope}
\draw (190.000000,27.000000) -- (190.000000,13.500000);
\begin{scope}
\draw (187.878680, 24.878680) -- (192.121320, 29.121320);
\draw (187.878680, 29.121320) -- (192.121320, 24.878680);
\end{scope}
\begin{scope}
\draw (187.878680, 11.378680) -- (192.121320, 15.621320);
\draw (187.878680, 15.621320) -- (192.121320, 11.378680);
\end{scope}
\draw (207.000000,40.500000) -- (207.000000,0.000000);
\begin{scope}
\draw (204.878680, 38.378680) -- (209.121320, 42.621320);
\draw (204.878680, 42.621320) -- (209.121320, 38.378680);
\end{scope}
\begin{scope}
\draw (204.878680, -2.121320) -- (209.121320, 2.121320);
\draw (204.878680, 2.121320) -- (209.121320, -2.121320);
\end{scope}
\end{tikzpicture}
    \caption{Example of a quantum Fourier transform circuit on four qubits as used in our calculations. The controlled operators $P_n = e^{-i\pi (\sigma^z - \mathds{1}) 2^{-n}}$ are phase gates with rotation angles of $2\pi/2^n$.}
    \label{fig:qft_circ}
\end{figure}
As shown in the previous subsection, when compiling random unitaries, the number of required parameters quickly makes compiling circuits with our available resources for more than a few qubits infeasible. However, practically relevant circuits usually have much more structure than random unitaries. We therefore also synthesise quantum Fourier transform (QFT)~\cite{nielsen2000quantum} circuits using various gate sets as examples of unitaries closer to real-world applications. \Cref{fig:qft_circ} shows the target circuit of a 4-qubit QFT as an example. We re-express it using the established gate set $\mathcal{L}_\mathrm{allrot}$, as well as
\begin{multline}
    \mathcal{L}_\mathrm{NNrot} = \{R^\sigma_k, C_{\ell}[R^\sigma_k] \:|\: \sigma \in \{x, y, z\}, \\ k, \ell \in [N] \text{ and } |k - \ell| = 1\}
\end{multline}
which contains all local single-qubit rotations and local rotations controlled by nearest neighbours in an open linear chain topology. To test the ability of our algorithms to work with a much more restricted gate set, we furthermore synthesise QFT circuits using the set
\begin{multline}
    \mathcal{L}_\mathrm{\textsc{swap}} = \{R^\sigma_k, E_{k, \ell} \:|\: \sigma \in \{x, y, z\},\\ k, \ell \in [N] \text{ and } |k - \ell| = 1\}
\end{multline}
which, in addition to local rotations, contains the parameterised \textsc{swap} gate introduced in \cref{eq:param_swap} between neighbouring qubits as the only entangling operator. None of these gate sets contain Hadamard or controlled phase gates -- which constitute the majority of the gates in the canonical circuit -- making this a suitable benchmarking synthesis task.

\begin{figure}[!htb]
    \input{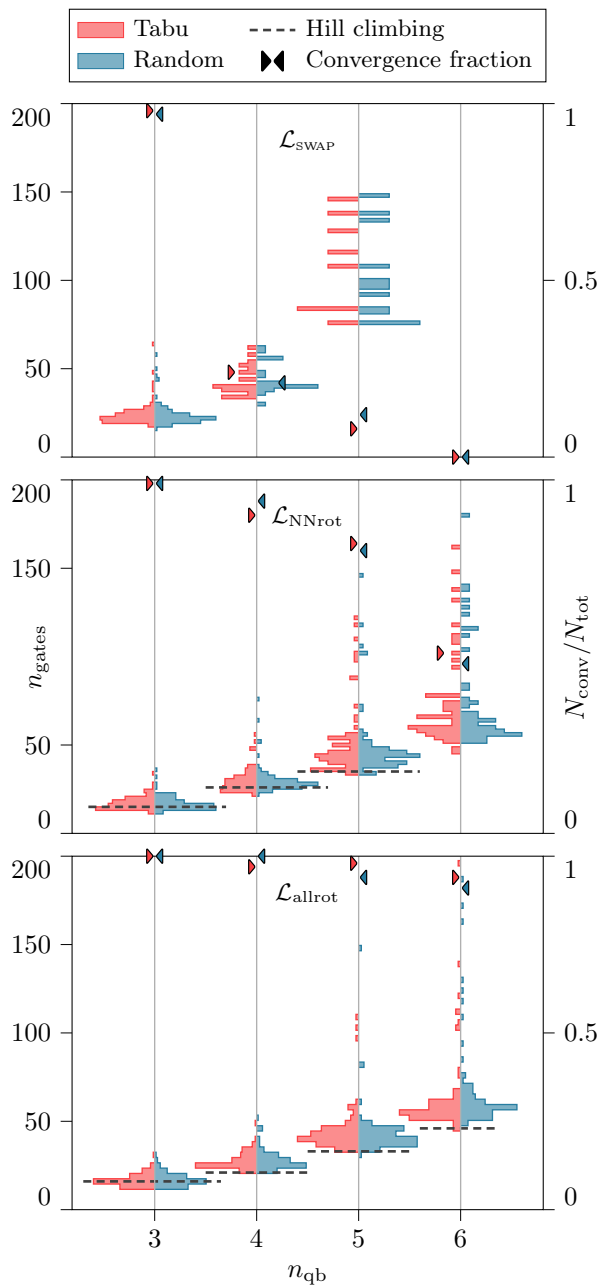}
    \caption{Number of gates and fraction of converged calculations for synthesis of a QFT circuit for $n_\mathrm{qb} = 3\ldots6$ qubits and different gate sets $\mathcal{L}_\mathrm{\textsc{swap}}$, $\mathcal{L}_\mathrm{NNrot}$, and $\mathcal{L}_\mathrm{allrot}$. Shaded graphs are histograms of the number of gates in the resultant circuit, left, red \ref{leg:tabu} for tabu search, right, blue \ref{leg:rand} for random search, centred around their respective number of qubits. Each calculation was run 100 times with identical parameters as listed in the text and appendix. Red and blue carets indicate the fraction of converged calculations for each instance according to the right scale. Separation from the central line is for visual purposes only. Grey dashed lines \ref{leg:hill}, if present, indicate that the hill climbing algorithm converged for a particular instance and shows the number of qubits in the final circuit it produced.}
    \label{fig:qft_gates}
\end{figure}

For demonstration purposes, we tried synthesising QFT circuits for 3 to 6 qubits, all mentioned gate sets, and all discussed circuit structure generation algorithms, 100 times each. The target energy for the used cost function was $\expval{H_\mathrm{sum}} \leq 10^{-8}$. The results are shown in \cref{fig:qft_gates}. In addition to the outcomes shown in \cref{fig:qft_gates}, we furthermore performed 10 calculations each for $n_\mathrm{qb} = 7\ldots 10$ qubits using tabu search and the $\mathcal{L}_\mathrm{allrot}$ gate set. These resulted in the convergence of 8, 7, 5, and 1 calculations, respectively, and minimum gate counts of 72, 93, 114, and 138.

We found that the \emph{hill climbing} algorithm can perform well for some problems, especially when the available gate set has high expressibility as is the case for $\mathcal{L}_\mathrm{allrot}$ and $\mathcal{L}_\mathrm{NNrot}$. In these cases, hill climbing found circuits with gate counts close to the lower bound of all algorithms we investigated, albeit at a higher computational cost than tabu and random search. However, more restricted gate sets like $\mathcal{L}_\mathrm{\textsc{swap}}$ severely hamper its ability to find solutions. Being the only deterministic algorithm we presented, circuits it fails to synthesise cannot be helped by re-running the procedure.

The probabilistic schemes of tabu search and random search, on the other hand, produce different outcomes for every run. We found that using a fully connected gate set consistently yields a higher probability for convergence than restricting the interactions to neighbouring qubits in a linear chain, despite using an initial circuit of \textsc{swap}s to counteract connectivity constraints. Calculations starting from an empty circuit (not plotted) show even lower convergence rates. Further decreasing the expressibility of the available gates by using $\mathcal{L}_\mathrm{\textsc{swap}}$ as the gate set sees another significant drop in the relative number of converged calculations for $n_\mathrm{qb} \geq 4$, indicating that all of our algorithms struggle to find solutions when they are greatly restricted in the choice of gates they can add.

Comparing our approaches of \emph{tabu} and \emph{random search}, we find -- as for random unitaries -- no significant difference in the number of gates in the resulting circuit or the probability convergence, despite tabu search trying to remember and avoid unsuccessful circuit modifications. It is to be expected for this mechanism not to have a noticeable impact when using libraries containing many gates, like $\mathcal{L}_\mathrm{allrot}$, because the neighbourhood -- i.e. the circuits which can be produced by adding a single gate from the library at a any position -- is much larger than the number of additions the algorithm can reasonably try during the iteration. For example, on 5 qubits with an existing circuit containing 20 gates, the $\mathcal{L}_\mathrm{allrot}$ gate set produces several hundred potential moves, of which our algorithms typically explore $\sim 20$. However, even when using the $\mathcal{L}_\mathrm{\textsc{swap}}$ library with only very few gates in it, which thus produces a smaller neighbourhood for each circuit, we found no significant difference between the two approaches. This observation is independent of how many iterations the \emph{tabu} moves are remembered for.

\subsection{\texorpdfstring{$n$}{n}-qubit Toffoli} \label{sec:toffoli}
A potentially difficult operator to synthesise from only two-qubit gates is the $n$-qubit Toffoli gate, i.e. a Pauli-$X$ gate with $n-1$ controls. The difficulty lies in the fact that it only acts on a very small subspace of $\mathcal{H}$, which is not straightforward to exclusively address using gates acting on much larger spaces of $\mathcal{H}$. Additionally, while it is relatively straightforward to detect whether a given circuit can act exclusively in the given subspace, finding a measure indicating that a given circuit is \emph{close} to having this property proves difficult. If this kind of measure were found, it could guide the compilation process in the right direction. Unfortunately, our used cost functions do not contain such information.

In our implementation, because a large portion of all possible input states must remain unchanged by the circuit, adding any small number of gates will likely result in their parameters being tuned to zero during optimisation, as that matches the correct action on most input states. The algorithm then subsequently removes the gates with vanishing parameters, leading to no progress being made.

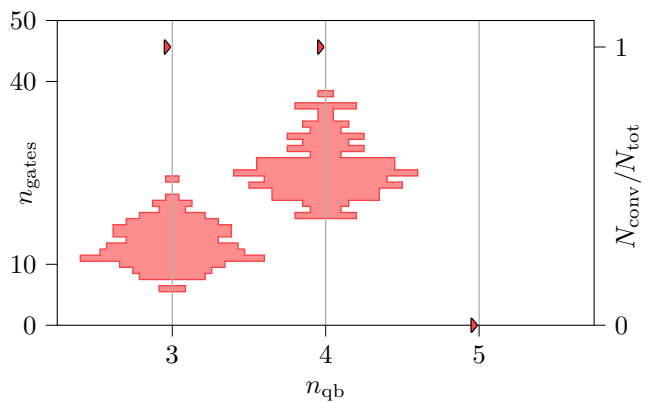
\begin{figure}[b]
\begin{tikzpicture}

\definecolor{darkgray176}{RGB}{176,176,176}
\definecolor{lightgray204}{RGB}{204,204,204}
\definecolor{tomato2496568}{RGB}{249,65,68}

\begin{axis}[
axis on top,
grid style={line width=.5pt},
height=16em,
legend cell align={left},
legend style={fill opacity=0.8, draw opacity=1, text opacity=1, draw=lightgray204},
tick align=outside,
tick pos=left,
xtick={3,4,5},
ytick={0,10,40,50},
ylabel={$n_\mathrm{gates}$},
unbounded coords=jump,
width=\columnwidth,
x grid style={darkgray176},
xlabel={\(\displaystyle n_\mathrm{qb}\)},
xmajorgrids,
xmin=2.25, xmax=5.75,
xtick style={color=black},
y grid style={darkgray176},
ymin=0, ymax=50,
ytick style={color=black},
y label style={yshift=-1.8em},
]
\path [fill=tomato2496568, fill opacity=0.6]
(axis cs:3,5.5)
--(axis cs:3,5.5)
--(axis cs:3,5.5)
--(axis cs:3,5.5)
--cycle;
\path [fill=tomato2496568, fill opacity=0.6]
(axis cs:3,5.5)
--(axis cs:3,5.5)
--(axis cs:2.91428571428571,5.5)
--(axis cs:2.91428571428571,6.5)
--(axis cs:3,6.5)
--(axis cs:3,6.5)
--(axis cs:3,6.5)
--(axis cs:3.08571428571429,6.5)
--(axis cs:3.08571428571429,5.5)
--(axis cs:3,5.5)
--cycle;
\path [fill=tomato2496568, fill opacity=0.6]
(axis cs:3,7.5)
--(axis cs:3,7.5)
--(axis cs:2.78571428571429,7.5)
--(axis cs:2.78571428571429,8.5)
--(axis cs:2.74285714285714,8.5)
--(axis cs:2.74285714285714,9.5)
--(axis cs:2.65714285714286,9.5)
--(axis cs:2.65714285714286,10.5)
--(axis cs:2.4,10.5)
--(axis cs:2.4,11.5)
--(axis cs:2.52857142857143,11.5)
--(axis cs:2.52857142857143,12.5)
--(axis cs:2.57142857142857,12.5)
--(axis cs:2.57142857142857,13.5)
--(axis cs:2.7,13.5)
--(axis cs:2.7,14.5)
--(axis cs:2.61428571428571,14.5)
--(axis cs:2.61428571428571,16.5)
--(axis cs:2.7,16.5)
--(axis cs:2.7,17.5)
--(axis cs:2.78571428571429,17.5)
--(axis cs:2.78571428571429,18.5)
--(axis cs:2.91428571428571,18.5)
--(axis cs:2.91428571428571,19.5)
--(axis cs:2.87142857142857,19.5)
--(axis cs:2.87142857142857,20.5)
--(axis cs:2.95714285714286,20.5)
--(axis cs:2.95714285714286,21.5)
--(axis cs:3,21.5)
--(axis cs:3,21.5)
--(axis cs:3,21.5)
--(axis cs:3.04285714285714,21.5)
--(axis cs:3.04285714285714,20.5)
--(axis cs:3.12857142857143,20.5)
--(axis cs:3.12857142857143,19.5)
--(axis cs:3.08571428571429,19.5)
--(axis cs:3.08571428571429,18.5)
--(axis cs:3.21428571428571,18.5)
--(axis cs:3.21428571428571,17.5)
--(axis cs:3.3,17.5)
--(axis cs:3.3,16.5)
--(axis cs:3.38571428571429,16.5)
--(axis cs:3.38571428571429,14.5)
--(axis cs:3.3,14.5)
--(axis cs:3.3,13.5)
--(axis cs:3.42857142857143,13.5)
--(axis cs:3.42857142857143,12.5)
--(axis cs:3.47142857142857,12.5)
--(axis cs:3.47142857142857,11.5)
--(axis cs:3.6,11.5)
--(axis cs:3.6,10.5)
--(axis cs:3.34285714285714,10.5)
--(axis cs:3.34285714285714,9.5)
--(axis cs:3.25714285714286,9.5)
--(axis cs:3.25714285714286,8.5)
--(axis cs:3.21428571428571,8.5)
--(axis cs:3.21428571428571,7.5)
--(axis cs:3,7.5)
--cycle;
\path [fill=tomato2496568, fill opacity=0.6]
(axis cs:3,23.5)
--(axis cs:3,23.5)
--(axis cs:2.95714285714286,23.5)
--(axis cs:2.95714285714286,24.5)
--(axis cs:3,24.5)
--(axis cs:3,24.5)
--(axis cs:3,24.5)
--(axis cs:3.04285714285714,24.5)
--(axis cs:3.04285714285714,23.5)
--(axis cs:3,23.5)
--cycle;
\path [fill=tomato2496568, fill opacity=0.6]
(axis cs:3,25.5)
--(axis cs:3,25.5)
--(axis cs:3,25.5)
--(axis cs:3,25.5)
--cycle;
\path [fill=tomato2496568, fill opacity=0.6]
(axis cs:3,25.5)
--(axis cs:3,25.5)
--(axis cs:3,25.5)
--(axis cs:3,25.5)
--cycle;

\path [fill=tomato2496568, fill opacity=0.6]
(axis cs:4,5.5)
--(axis cs:4,5.5)
--(axis cs:4,5.5)
--(axis cs:4,5.5)
--cycle;
\path [fill=tomato2496568, fill opacity=0.6]
(axis cs:4,5.5)
--(axis cs:4,5.5)
--(axis cs:4,5.5)
--(axis cs:4,5.5)
--cycle;
\path [fill=tomato2496568, fill opacity=0.6]
(axis cs:4,5.5)
--(axis cs:4,5.5)
--(axis cs:4,5.5)
--(axis cs:4,5.5)
--cycle;
\path [fill=tomato2496568, fill opacity=0.6]
(axis cs:4,17.5)
--(axis cs:4,17.5)
--(axis cs:3.8,17.5)
--(axis cs:3.8,18.5)
--(axis cs:3.9,18.5)
--(axis cs:3.9,19.5)
--(axis cs:3.85,19.5)
--(axis cs:3.85,20.5)
--(axis cs:3.65,20.5)
--(axis cs:3.65,22.5)
--(axis cs:3.5,22.5)
--(axis cs:3.5,23.5)
--(axis cs:3.6,23.5)
--(axis cs:3.6,24.5)
--(axis cs:3.4,24.5)
--(axis cs:3.4,25.5)
--(axis cs:3.55,25.5)
--(axis cs:3.55,27.5)
--(axis cs:3.9,27.5)
--(axis cs:3.9,28.5)
--(axis cs:3.75,28.5)
--(axis cs:3.75,29.5)
--(axis cs:3.85,29.5)
--(axis cs:3.85,30.5)
--(axis cs:3.75,30.5)
--(axis cs:3.75,31.5)
--(axis cs:3.9,31.5)
--(axis cs:3.9,32.5)
--(axis cs:3.85,32.5)
--(axis cs:3.85,33.5)
--(axis cs:3.95,33.5)
--(axis cs:3.95,35.5)
--(axis cs:3.8,35.5)
--(axis cs:3.8,36.5)
--(axis cs:4,36.5)
--(axis cs:4,36.5)
--(axis cs:4,36.5)
--(axis cs:4.2,36.5)
--(axis cs:4.2,35.5)
--(axis cs:4.05,35.5)
--(axis cs:4.05,33.5)
--(axis cs:4.15,33.5)
--(axis cs:4.15,32.5)
--(axis cs:4.1,32.5)
--(axis cs:4.1,31.5)
--(axis cs:4.25,31.5)
--(axis cs:4.25,30.5)
--(axis cs:4.15,30.5)
--(axis cs:4.15,29.5)
--(axis cs:4.25,29.5)
--(axis cs:4.25,28.5)
--(axis cs:4.1,28.5)
--(axis cs:4.1,27.5)
--(axis cs:4.45,27.5)
--(axis cs:4.45,25.5)
--(axis cs:4.6,25.5)
--(axis cs:4.6,24.5)
--(axis cs:4.4,24.5)
--(axis cs:4.4,23.5)
--(axis cs:4.5,23.5)
--(axis cs:4.5,22.5)
--(axis cs:4.35,22.5)
--(axis cs:4.35,20.5)
--(axis cs:4.15,20.5)
--(axis cs:4.15,19.5)
--(axis cs:4.1,19.5)
--(axis cs:4.1,18.5)
--(axis cs:4.2,18.5)
--(axis cs:4.2,17.5)
--(axis cs:4,17.5)
--cycle;
\path [fill=tomato2496568, fill opacity=0.6]
(axis cs:4,37.5)
--(axis cs:4,37.5)
--(axis cs:3.95,37.5)
--(axis cs:3.95,38.5)
--(axis cs:4,38.5)
--(axis cs:4,38.5)
--(axis cs:4,38.5)
--(axis cs:4.05,38.5)
--(axis cs:4.05,37.5)
--(axis cs:4,37.5)
--cycle;
\path [fill=tomato2496568, fill opacity=0.6]
(axis cs:4,70.5)
--(axis cs:4,70.5)
--(axis cs:3.95,70.5)
--(axis cs:3.95,71.5)
--(axis cs:4,71.5)
--(axis cs:4,71.5)
--(axis cs:4,71.5)
--(axis cs:4.05,71.5)
--(axis cs:4.05,70.5)
--(axis cs:4,70.5)
--cycle;

\addplot [line width=0.5pt, tomato2496568, forget plot]
table {%
3 5.5
nan nan
3 5.5
3.08571428571429 5.5
3.08571428571429 6.5
3 6.5
nan nan
3 7.5
3.21428571428571 7.5
3.21428571428571 8.5
3.25714285714286 8.5
3.25714285714286 9.5
3.34285714285714 9.5
3.34285714285714 10.5
3.6 10.5
3.6 11.5
3.47142857142857 11.5
3.47142857142857 12.5
3.42857142857143 12.5
3.42857142857143 13.5
3.3 13.5
3.3 14.5
3.38571428571429 14.5
3.38571428571429 16.5
3.3 16.5
3.3 17.5
3.21428571428571 17.5
3.21428571428571 18.5
3.08571428571429 18.5
3.08571428571429 19.5
3.12857142857143 19.5
3.12857142857143 20.5
3.04285714285714 20.5
3.04285714285714 21.5
3 21.5
nan nan
3 23.5
3.04285714285714 23.5
3.04285714285714 24.5
3 24.5
nan nan
3 25.5
nan nan
3 25.5
};
\addplot [line width=0.5pt, tomato2496568, forget plot]
table {%
3 5.5
nan nan
3 5.5
2.91428571428571 5.5
2.91428571428571 6.5
3 6.5
nan nan
3 7.5
2.78571428571429 7.5
2.78571428571429 8.5
2.74285714285714 8.5
2.74285714285714 9.5
2.65714285714286 9.5
2.65714285714286 10.5
2.4 10.5
2.4 11.5
2.52857142857143 11.5
2.52857142857143 12.5
2.57142857142857 12.5
2.57142857142857 13.5
2.7 13.5
2.7 14.5
2.61428571428571 14.5
2.61428571428571 16.5
2.7 16.5
2.7 17.5
2.78571428571429 17.5
2.78571428571429 18.5
2.91428571428571 18.5
2.91428571428571 19.5
2.87142857142857 19.5
2.87142857142857 20.5
2.95714285714286 20.5
2.95714285714286 21.5
3 21.5
nan nan
3 23.5
2.95714285714286 23.5
2.95714285714286 24.5
3 24.5
nan nan
3 25.5
nan nan
3 25.5
};
\addplot [line width=0.5pt, tomato2496568, forget plot]
table {%
4 5.5
nan nan
4 5.5
nan nan
4 5.5
nan nan
4 17.5
4.2 17.5
4.2 18.5
4.1 18.5
4.1 19.5
4.15 19.5
4.15 20.5
4.35 20.5
4.35 22.5
4.5 22.5
4.5 23.5
4.4 23.5
4.4 24.5
4.6 24.5
4.6 25.5
4.45 25.5
4.45 27.5
4.1 27.5
4.1 28.5
4.25 28.5
4.25 29.5
4.15 29.5
4.15 30.5
4.25 30.5
4.25 31.5
4.1 31.5
4.1 32.5
4.15 32.5
4.15 33.5
4.05 33.5
4.05 35.5
4.2 35.5
4.2 36.5
4 36.5
nan nan
4 37.5
4.05 37.5
4.05 38.5
4 38.5
nan nan
4 70.5
4.05 70.5
4.05 71.5
4 71.5
};
\addplot [line width=0.5pt, tomato2496568, forget plot]
table {%
4 5.5
nan nan
4 5.5
nan nan
4 5.5
nan nan
4 17.5
3.8 17.5
3.8 18.5
3.9 18.5
3.9 19.5
3.85 19.5
3.85 20.5
3.65 20.5
3.65 22.5
3.5 22.5
3.5 23.5
3.6 23.5
3.6 24.5
3.4 24.5
3.4 25.5
3.55 25.5
3.55 27.5
3.9 27.5
3.9 28.5
3.75 28.5
3.75 29.5
3.85 29.5
3.85 30.5
3.75 30.5
3.75 31.5
3.9 31.5
3.9 32.5
3.85 32.5
3.85 33.5
3.95 33.5
3.95 35.5
3.8 35.5
3.8 36.5
4 36.5
nan nan
4 37.5
3.95 37.5
3.95 38.5
4 38.5
nan nan
4 70.5
3.95 70.5
3.95 71.5
4 71.5
};
\end{axis}

\begin{axis}[
axis on top,
axis y line=right,
grid style={line width=.5pt},
height=15em,
tick align=outside,
unbounded coords=jump,
width=\columnwidth,
x grid style={darkgray176},
xmin=2.25, xmax=5.75,
xtick pos=left,
xtick style={color=black},
y grid style={darkgray176},
ymin=0, ymax=1,
ytick pos=right,
ytick style={color=black},
yticklabel style={anchor=west},
hide x axis,
axis line style={draw=none},
ylabel={$N_\mathrm{conv}/N_\mathrm{tot}$},
y label style={yshift=1.2em},
ytick={0,1},
]
\addplot [semithick, black, mark=caretleft, mark size=6, mark options={solid,fill=tomato2496568}]
table {%
3 1
};
\addplot [semithick, black, mark=caretleft, mark size=6, mark options={solid,fill=tomato2496568}]
table {%
4 1
};
\addplot [semithick, black, mark=caretleft, mark size=6, mark options={solid,fill=tomato2496568}]
table {%
5 0
};
\end{axis}
\pgfresetboundingbox
\useasboundingbox
(-.7,-1.1) rectangle (7.8,4.3);


\end{tikzpicture}
    \caption{Histograms of the numbers of gates required and convergence fractions when synthesising an $n_\mathrm{qb}$-qubit Toffoli using the $\mathcal{L}_\mathrm{allrot}$ gate set and tabu search. Histograms are centred around the corresponding number of qubits, red carets indicate how many of the started calculations converged. Histogram data for 5 qubits is missing because none of the calculations succeeded.}
    \label{fig:toffoli_gates}
\end{figure}

\Cref{fig:toffoli_gates} shows the results for our synthesis calculations to generate $n$-qubit Toffoli gates for $n = 3, 4, 5$ using the $\mathcal{L}_\mathrm{allrot}$ gate library, each 100 times. For the smaller cases of $n=3$ and $4$ our algorithm can find correct circuits reliably. In these cases the number of simultaneously added gates is large enough to overcome the previously discussed limitation. This is helped by the fact that the number of states on which the Toffoli acts like the identity operator is not overwhelmingly bigger than the number of states on which it acts nontrivially. Therefore, in these cases, we succeed in synthesising an appropriate circuit virtually all of the time. For $n \geq 5$, on the other hand, our algorithm was not able to find any solutions at all.

As an attempt to inject some prior knowledge into the algorithm, we also performed 5-qubit Toffoli gate synthesis starting the iteration from one of the successful 4-qubit Toffoli synthesis results. With this `warm start' technique, 80 out of 100 calculations converged. While such an assisted start deviates from the strict \textit{ab initio} framework -- even more so than the \textsc{swap} network used in some of our calculations -- and requires a specific incremental structure of the target circuit, it can still be a very useful resource for some tasks.

\section{Discussion and Outlook} \label{sec:discussion}

In this work we have combined a variant of the Hilbert-Schmidt test~\cite{khatri2019quantumassisted} with artificial Hamiltonians $H_\mathrm{sum}$ and $H_\mathrm{proj}$ similar to Ref.~\cite{jones2022robust} to construct cost functions representing the closeness of a unitary to a dynamically created circuit, where the cost can in principle be evaluated on a quantum computer. However, we only employed emulators of such quantum hardware, and have thus circumvented some practical challenges like shot noise and barren plateaus, whose impact on the performance on the presented scheme remains to be investigated. We presented three different algorithms which use this cost function to dynamically construct quantum circuits replicating the action of a given unitary from the ground up and demonstrated their performance on various synthesis tasks.

Our results suggest that the presented algorithms are able to generate circuit representations for dense random unitaries with gate counts close to what we would expect to be optimal, based on the degrees of freedom in such a unitary. For block-diagonal random unitaries we were furthermore able to show that generating a circuit whose closeness to the target is only judged within a restricted subspace greatly reduces both the synthesis resource requirements and the gate count in the resulting circuit. This can be important when time evolving Hamiltonians with such a block-diagonal structure, as is usually the case in quantum chemistry.

None of the cases we numerically investigated showed a significant difference between \emph{random search} and \emph{tabu search}. This strongly suggests that our simple attempt at guiding the search through the circuit structure space more efficiently than random moves is not sophisticated enough to yield any practical advantage. We note that the implementation of tabu search used here, only incorporates its most basic aspect of short-term memory. Its performance could potentially be improved by including more elaborate concepts such as intermediate-term and long-term memory~\cite{glover1990tabu}.

The presented results for synthesising quantum Fourier transform circuits using various gate sets show that for small numbers of qubits and highly expressive sets of gates, the hill climbing method can consistently produce circuits with very few gates, almost always close to the minimum number we found for any method. However, if the method gets stuck in a local minimum or proceeds too slowly due to the increasing size of the neighbourhood, there are no provisions in our presented framework to overcome these problems. Tabu and random search produced accurate results even for a very restricted gate set on a small number of qubits, but scaled unfavourably when increasing the number of qubits in these cases.

Finally, the attempts at synthesising an $n$-qubit Toffoli gate clearly showed the limitations of our method. Due to the properties of the cost functions, as discussed in \cref{sec:toffoli}, we were only able to generate circuits for Toffoli gates on 5 qubits by assisting the algorithm with previously generated knowledge.

We emphasise that there is significant value in the ability to synthesise even small unitaries, since such compiled functions can be used as components of larger algorithms. For example, in grid-based chemistry, although the total number of computational qubits may be in the many thousands, the QFT used to move between real-space and momentum-space is local to each particle's representation, often acting on only around 20 qubits~\cite{chan2022grid}. Indeed, even the ability to compile a multi-qubit gate involving 3 or 4 qubits into a compact set of 1- and 2-qubit gates can be valuable. Therefore the significance of the techniques described in this paper does not depend on their ability to scale directly to circuit sizes that might be considered `post-classical' (\raisebox{.2ex}{\scriptsize $\gtrsim$} 50 qubits). Nevertheless it is of course interesting to reflect on the prospects for such large scale circuit synthesis and (re-)compilations. Our results suggest that the methods will require significant further development for any such task to be realistic. We now remark on a few such possibilities.

Firstly, note that the algorithms used for circuit structure modifications are largely independent from the parameter optimisation routine, except for reusing information in the quantum metric tensor. Therefore, if a different method for optimising the parameters proves more suitable, it can be straightforwardly substituted for the imaginary time evolution used in the present paper. One promising candidate is introduced in Ref.~\cite{boyd2022covar} named CoVaR, where an eigenstate of the system is prepared by a root-finding algorithm similar to Newton's method. By tweaking the spectrum of our synthesis Hamiltonians such that only product states are eigenstates, this method could be used to find the appropriate parameters in each iteration.

Secondly, the gate sets need not consist only of unitary operators. Instead, it would be possible to also include ancilla qubits on which intermediate measurements may be performed, whose outcomes can become part of the cost function. In this case, imaginary time evolution must be adapted to find the correct descent direction \cite{koczor2019natural}. 

Thirdly, there are also possible enhancements to the Hamiltonians used to generate our cost functions. As briefly mentioned in \cref{sec:cost_functions}, instead of the relatively straightforward $H_\mathrm{proj}$ and $H_\mathrm{sum}$, other properties of the solution may be included to judge how suitable a particular outcome is, such as the desired entanglement via a witness, or the conservation of symmetries in the problem.

Fourthly and perhaps most challengingly, an intriguing direction of research is to explore methods for introducing circuit variants that are more nuanced than the Darwinian `random variation, non-random selection' employed in this paper. While enhancements to the tabu search introduced here may form part of the solution, another way forward might be to explore techniques that use information from the output state to deduce which modifications to the circuit are most likely to result in a reduction of the cost function.

Finally we remark that the presented methods to construct circuits \textit{ab initio} can be used not only to express a desired unitary using various target gate sets, but also as a variational quantum eigensolver to prepare the ground state of some physical Hamiltonian. For this task, the evaluation of the cost function is straightforwardly replaced by the Hamiltonian of interest. This idea, among other applications, is explored in the sister paper to the present work~\cite{gustiani2022variational}.

\section*{Acknowledgements}
    The authors would like to thank Tyson Jones and Bálint Koczor for useful discussions and feedback on the manuscript. 
    CG and SCB acknowledge financial support from EPSRC Hub
grants under agreement No.~EP/T001062/1, from the IARPA funded LogiQ
project, and the EU flagship AQTION project.

The authors acknowledge the use of the University of Oxford Advanced Research Computing (ARC) facility in carrying out this work {(\small \doi{10.5281/zenodo.22558})}.

\interlinepenalty=8000
\bibliography{references}
\newpage
\appendix

\section{Algorithmic details}

\subsection{Global hyperparameters} \label{sec:hyperparams}
In order to make the algorithms easier to read, we globally define some hyperparameters in \cref{tbl:hyperparams}. They stay constant during the whole synthesis process and can be used to tweak some properties of the algorithms.

\begin{table}[htb]
    \caption{Hyperparameters used in the pseudocode of our subroutines, collected here for more concise descriptions of the algorithms.}
    \label{tbl:hyperparams}
    \vspace{1.5ex}
    \footnotesize
    \setlength\tabcolsep{.9em}
    \begin{tabularx}{\linewidth}{llX}\toprule
        {Param} & {Default} & {Use} \\ \midrule
        $\tilde{H}$ & $H_\mathrm{sum}$ & The synthesis Hamiltonian in the augmented space. In our case either $H_\mathrm{sum}$ or $H_\mathrm{proj}$.\\
        $\delta_\mathrm{abs}$ & $10^{-5}$ & Threshold for the absolute ener\-gy change per iteration regarded as constant during the parameter optimisation.\\
        $\delta_\mathrm{rel}$ & $10^{-3}$ & Limit for the relative energy change per iteration considered constant during the parameter optimisation.\\
        $k_\mathrm{max,opt}$ & 500 & Maximum number of iterations for parameter optimisation.\\
        $n_\mathrm{conv}$ & 5 & Number of times the convergence criterion must be fulfilled in parameter optimisation to be considered converged.\\
        $\kappa$ & 1.4 & Factor by which the step size $\lambda$ of the parameter optimisation is in- or decreased while searching along the gradient direction.\\
        $\lambda_0$ & 0.05 & Initial step size in the parameter optimisation routine.\\
        $k_\mathrm{max}$ & 10\,000 & Maximum number of iterations for circuit modifications.\\
        $E_\mathrm{conv}$ & $10^{-8}$ & Energy at which the calculation is considered converged.\\
        $N_\mathrm{moves}$ & 30 & Number of gates added in a single circuit modification iteration for random and tabu search.\\
        $N_\mathrm{samp}$ & 10 & Number of times a circuit is modified in a single iteration until the best result is picked for the next iteration in random search.\\
        $\mathcal{L}$ & $\mathcal{L}_\mathrm{allrot}$ & The library to draw new gates from.\\
        $\varepsilon_\mathrm{QMT}$ & $10^{-3}$ & Small quantity to detect linearly dependent rows in the QMT.\\
        $\varepsilon_\mathrm{param}$ & $\varepsilon_\mathrm{remove}$ & Threshold for the magnitude of parameters below which the corresponding gate is removed.\\
        $\varepsilon_\mathrm{remove}$ & $\frac{E_k - E_{k-1}}{50}$ & Energy increase considered acceptable for the removal of unnecessary gates.\\\bottomrule
    \end{tabularx}
\end{table}

\newpage
\subsection{Pseudocode} \label{sec:pseudocode}

To minimise clutter in the main text, we collect pseudocode for most of our described routines in this appendix for the interested reader. In contrast to the main text, where we write the cost function as $\expval*{\tilde{H}}$, in pseudocode to explicitly denote which circuit is applied, we use the notation
\begin{equation}
    \mathcal{E}(\mathcal{C}(\vec{\theta})) =  \expval*{\tilde{H}}{\psi_1}
\end{equation}
with
\begin{equation}
    \ket{\psi_1} = P^\dagger\mathcal{C}^\dagger(\vec{\theta}) U P (\ket{0}_{\mathcal{H}}\otimes\ket{0}_{\mathcal{H}'})
\end{equation}
 where operator $P$ prepares Bell pairs as depicted on the left side of \cref{fig:fullspace_setup}.

\begin{algorithm}[H] \small
    \setstretch{1.1}
    \begin{algorithmic}
        \Function{OptimiseParameters}{$\mathcal{C}$, $\vec{\theta}$}
            \State $E_0 \gets \mathcal{E}(\mathcal{C}(\vec{\theta}))$,\quad $k_\mathrm{conv} \gets 0$
            \For{$k \gets 1\,..\,k_\mathrm{max,opt}$}
                \State $A_{ij} \gets \Re(\braket{\partial_i \psi}{\partial_j \psi} - \braket{\partial_i \psi}{\psi}\braket{\psi}{\partial_j \psi}) \;\forall\, i, j$
                \State $B_i \gets -\mel*{\partial_i \psi}{\hat{H}}{\psi} \;\forall\, i$
                \State $\vec{\Delta} \gets \textproc{Regularise}(\mat{A})^{-1}\,\vec{B}$
                \State $\lambda \gets \textproc{ChooseLambda}(\lambda_0, \mathcal{C}, \vec{\theta}, \vec{\Delta})$
                \State $\vec{\theta} \gets \vec{\theta} - \lambda \vec{\Delta}$
                \State $E_k \gets \mathcal{E}(\mathcal{C}(\vec{\theta}))$
                \If{$E_k - E_{k-1} < \delta_\mathrm{abs}$ \Or $\frac{E_k - E_{k-1}}{E_k} < \delta_\mathrm{rel}$}
                    \State $k_\mathrm{conv} \gets k_\mathrm{conv} + 1$
                \Else
                    \State $k_\mathrm{conv} \gets 0$
                \EndIf
                \If{$k_\mathrm{conv} = n_\mathrm{conv}$}
                    \Return $\vec{\theta}$
                \EndIf
            \EndFor
            \State \Return \textbf{fail} \Comment{No convergence at max iterations}
        \EndFunction\\
        \Function{ChooseLambda}{$\lambda$, $\mathcal{C}$, $\vec{\theta}$, $\vec{\Delta}$}
            \State $\lambda^- \gets \lambda / \kappa, \quad \lambda^+ \gets \lambda \cdot \kappa, \quad E \gets \mathcal{E}(\mathcal{C}(\vec{\theta} - \lambda \vec{\Delta}))$
            \State $E^+ \gets \mathcal{E}(\mathcal{C}(\vec{\theta} - \lambda^+ \vec{\Delta})), \quad E^- \gets \mathcal{E}(\mathcal{C}(\vec{\theta} - \lambda^- \vec{\Delta}))$
            \If{$E^- > E < E^+$ \Or $\lambda \leq \lambda_\text{min}$}
                \State \Return $\lambda$ \Comment{no improvement either side}
            \EndIf
            \If{$E^+ < E^-$}
                \State \Return $\textproc{ChooseLambda}(\lambda^+, \mathcal{C}, \mathcal{\theta}, \vec{\Delta})$ \Comment{grow $\lambda$}
            \EndIf
            \State \Return $\textproc{ChooseLambda}(\lambda^-, \mathcal{C}, \mathcal{\theta}, \vec{\Delta})$ \Comment{shrink $\lambda$}
        \EndFunction
    \end{algorithmic}
    \caption{The routine we use to optimise the parameters $\vec{\theta}$ for a given circuit structure $\mathcal{C}$.}
    \label{algo:min_energy}
\end{algorithm}

\newpage

\begin{algorithm}[H] \small
    \setstretch{1.1}
    \begin{algorithmic}
        \Function{GenerateMoves}{$\mathcal{C}$}
            \State $\mathcal{M} \gets \{\}$
            \For{$k \gets 1\,..\,N_\text{qubits}$}
                \LineComment{Indices of gates touching qubit $k$}
                \State $\tilde{N} \gets (p\:|\:\mathcal{C}_p\text{ acts on qubit }k)$
                \State $G_\mathrm{left} \gets \mathds{1}$
                \ForAll{$m \in \tilde{N}$}
                    \State $\mathcal{M} \gets \mathcal{M} \cup \{(G, m)\:|\:G\in\mathcal{L}~\mathrm{and}~G \neq G_\mathrm{left}~$
                    \State $\hphantom{\mathcal{M} \gets \mathcal{M} \cup \{(G, m)\:|\:}\mathrm{and}~G \neq \mathcal{C}_m\}$
                    \State $G_\text{left} \gets \mathcal{C}_m$
                \EndFor
                \State $\mathcal{M} \gets \mathcal{M} \cup \{(G, \max(\tilde{N}) + 1)\:|\:G \in \mathcal{L}$
                \State $\hphantom{\mathcal{M} \gets \mathcal{M} \cup \{(G, \max(\tilde{N}) + 1)\:|\:}\mathrm{and}~G \neq G_\text{left}\}$
            \EndFor
            \State \Return $\mathcal{M}$
        \EndFunction\\
        
        \Function{ApplyMove}{$\mathcal{C}$, $\vec{\theta}$, $M$}
            \State $\mathcal{C} \gets (\mathcal{C}_0, \ldots, \mathcal{C}_{M_n - 1}, M_G, \mathcal{C}_{M_n}, \ldots)$
            \State $\vec{\theta} \gets (\theta_0, \ldots \theta_{M_n - 1}, 0, \theta_{M_n}, \ldots)$
            \State \Return $\mathcal{C}, \vec{\theta}$
        \EndFunction
    \end{algorithmic}
    \caption{Routine $\textproc{GenerateMoves}$ to generate all moves applicable to a circuit structure $\mathcal{C}$ not leading to obvious redundancies in the resulting circuit. The helper function $\textproc{ApplyMove}$ applies the move $M$ to the circuit structure $\mathcal{C}$ and parameter vector $\vec{\theta}$, and makes other routines easier to read.}
    \label{algo:moves}
\end{algorithm}

\begin{algorithm}[H] \small
    \setstretch{1.1}
    \begin{algorithmic}
        \Function{RandomSearch}{$\mathcal{C}^{(0)}$, $\vec{\theta}^{(0)}$}
            \State $E_0 \gets \mathcal{E}(\mathcal{C}^{(0)}(\vec{\theta}^{(0)}))$
            \For{$k \gets 1\,..\,k_\text{max}$}
                \LineComment $\mathcal{R}$ and $\tilde{\mathcal{R}}$ contain indices of newly added gates
                \State $E_k \gets E_{k-1}$, \quad $\mathcal{R} \gets \{\}$
                \State $\mathcal{C}^{(k)} \gets \mathcal{C}^{(k-1)}$, \quad $\vec{\theta}^{(k)} \gets \vec{\theta}^{(k-1)}$
                \For{$m \gets 1\,..\,N_\text{samp}$}
                    \State $\tilde{\mathcal{C}} \gets \mathcal{C}^{(k-1)}$,\quad $\tilde{\vec{\theta}} \gets \vec{\theta}^{(k-1)}$,\quad $\tilde{\mathcal{R}} \gets \{\}$
                    \For{$n \gets 1\,..\,N_\text{moves}$}
                        \State $M \gets \textbf{random element}$
                        \State $\hphantom{M \gets {}}\hspace{\algorithmicindent} \textbf{of } \textproc{GenerateMoves}(\tilde{\mathcal{C}})$
                        \State $\tilde{\mathcal{C}}, \tilde{\vec{\theta}} \gets \textproc{ApplyMove}(\tilde{\mathcal{C}}, \tilde{\vec{\theta}}, M)$
                        \State $\tilde{\mathcal{R}} \gets \{j\:|\:j\in\tilde{\mathcal{R}}~\mathrm{and}~j<M_n\}\cup\{M_n\}$
                \State $\hphantom{\mathcal{R} \gets \hspace{\algorithmicindent}}\cup\{j+1\:|\:j\in\tilde{\mathcal{R}}~\mathrm{and}~j\geq M_n\}$
                    \EndFor
                    \State $\tilde{\vec{\theta}} \gets \textproc{OptimiseParameters}(\tilde{\mathcal{C}}, \tilde{\vec{\theta}})$
                    \If{$\mathcal{E}(\tilde{\mathcal{C}}(\tilde{\vec{\theta}})) < E_k$}
                        \State $\mathcal{C}^{(k)} \gets \tilde{\mathcal{C}}$, \quad $\vec{\theta}^{(k)} \gets \tilde{\vec{\theta}}$
                        \State $E_k \gets \mathcal{E}(\tilde{\mathcal{C}}(\tilde{\vec{\theta}}))$,\quad $\mathcal{R}\gets\tilde{\mathcal{R}}$
                    \EndIf
                \EndFor
                \State $\mathcal{C}^{(k)}, \vec{\theta}^{(k)} \gets \textproc{Prune}(\mathcal{C}^{(k)}, \vec{\theta}^{(k)}, \mathcal{R})$
                \State $E_k \gets \mathcal{E}(\mathcal{C}^{(k)}(\vec{\theta}^{(k)}))$
                \If{$E_k < E_\text{conv}$}
                    \State \Return $\textproc{Prune}(\mathcal{C}^{(k)}, \vec{\theta}^{(k)}, \{0, \ldots, N_\mathrm{gates}(\mathcal{C}^{(k)})\})$
                \EndIf
            \EndFor
            \State \Return \textbf{fail} \Comment{No convergence at iteration limit}
        \EndFunction
    \end{algorithmic}
    \caption{A simplified version of the random search algorithm used to generate our results. It calls several subroutines from \cref{algo:moves,algo:min_energy,algo:prune}.}
    \label{algo:randomsearch}
\end{algorithm}

\newpage

\begin{algorithm}[H] \small
    \setstretch{1.1}
    \begin{algorithmic}
        \Function{Prune}{$\mathcal{C}$, $\vec{\theta}$, $\mathcal{R}$}
            \LineComment{Remove vanishing parameters}
            \State $\mathcal{D}\gets \{k\:|\:|\theta_k \bmod 2\pi| < \varepsilon_\mathrm{param}\}$
            \State $\mathcal{C}, \vec{\theta}, \mathcal{R} \gets \textproc{Delete}(\mathcal{C}, \vec{\theta}, \mathcal{R}, \mathcal{D})$
            \LineComment{Quantum metric tensor assisted removal}
            \State $\mathcal{D} \gets \{\}$, \quad $\mathcal{K} \gets \{\}$, \quad $\vec{\theta}^+ \gets \vec{0}$
            \State $A_{ij} \gets \Re(\braket{\partial_i \psi}{\partial_j \psi} - \braket{\partial_i \psi}{\psi}\braket{\psi}{\partial_j \psi}) \;\forall\, i, j$
            \For{$k \in \mathcal{R}$}
                \State $\mathcal{D} \gets \mathcal{D} \cup \{n\:|\:n>k~\text{and}$
                \State $\hphantom{\mathcal{D} \gets \{n\:|}\:\big|(A_{k,\cdot}^\intercal \cdot A_{n,\cdot}) - \|A_{k,\cdot}\|\,\|A_{n,\cdot}\| \,\big| < \varepsilon_\mathrm{QMT}\}$
                \ForAll{$n \in \mathcal{D}$}
                    \State $\mathcal{C}', \vec{\theta}' \gets \textproc{Delete}(\mathcal{C}, \vec{\theta}, \{\}, \{n\})$
                    \State $\theta_k' \gets \theta_k + \theta_n$
                    \If{$\mathcal{E}(\mathcal{C}'(\vec{\theta}')) > \mathcal{E}(\mathcal{C}(\mathcal{\theta})) + \varepsilon_\mathrm{remove}$}
                        \State $\mathcal{K} \gets \mathcal{K} \cup \{n\}$
                    \Else
                        \State $\vec{\theta}^+_k \gets \vec{\theta}^+_k + \theta_n$
                    \EndIf
                \EndFor
            \EndFor
            \State $\vec{\theta} \gets \vec{\theta} + \vec{\theta}^+$
            \State $\mathcal{C}, \vec{\theta}, \mathcal{R} \gets \textproc{Delete}(\mathcal{C}, \vec{\theta}, \mathcal{R}, \mathcal{D}\backslash\mathcal{K})$
            \LineComment{Trial and error removal}
            \While{$\mathcal{R} \neq \{\}$}
                \State $k \gets \max(\mathcal{R})$
                \State $\mathcal{R} \gets \mathcal{R} \backslash \{k\}$
                \State $\mathcal{C}' \gets (\mathcal{C}_0, \ldots \mathcal{C}_{k-1}, \mathcal{C}_{k+1}, \ldots)$
                \State $\vec{\theta}' \gets (\theta_0, \ldots \theta_{k-1}, \theta_{k+1}, \ldots)$
                \State $\vec{\theta}' \gets \textproc{OptimiseParameters}(\mathcal{C}', \vec{\theta}')$
                \If{$\mathcal{E}(\mathcal{C}'(\vec{\theta}')) < \mathcal{E}(\mathcal{C}(\vec{\theta})) + \varepsilon_\mathrm{remove}$}
                    \State $\mathcal{C} \gets \mathcal{C}'$
                    \State $\vec{\theta} \gets \vec{\theta}'$
                \EndIf
            \EndWhile
            \State \Return $\mathcal{C}$, $\vec{\theta}$
        \EndFunction\\
 
        \Function{Delete}{$\mathcal{C}$, $\vec{\theta}$, $\mathcal{R}$, $\mathcal{D}$}
            \For{$d \gets \textproc{InverseSorted}(\mathcal{D})$}
                \State $\mathcal{C} \gets (\mathcal{C}_0, \ldots \mathcal{C}_{d-1}, \mathcal{C}_{d+1}, \ldots)$
                \State $\vec{\theta} \gets (\theta_0, \ldots \theta_{d-1}, \theta_{d+1}, \ldots)$
                \State $\mathcal{R} \gets \{n\:|\:n\in\mathcal{R}~\mathrm{and}~n<d\}$
                \State $\hphantom{\mathcal{R} \gets \hspace{\algorithmicindent}}\cup\{n-1\:|\:n\in\mathcal{R}~\mathrm{and}~n>d\}$
            \EndFor
            \State \Return $\mathcal{C}$, $\vec{\theta}$, $\mathcal{R}$
        \EndFunction
    \end{algorithmic}
    \caption{Our routine to detect and remove unnecessary gates from a circuit. $\mathcal{C}$ and $\vec{\theta}$ are the circuit structure and current parameters, respectively, and $\mathcal{R}$ is a set of indices of gates that should be considered for deletion. The helper function $\textproc{Delete}$ removes the gates at the indices specified in $\mathcal{D}$ from the circuit and updates the indices in $\mathcal{R}$ accordingly.}
    \label{algo:prune}
\end{algorithm}

\begin{algorithm}[H] \small
    \setstretch{1.1}
    \begin{algorithmic}
        \Function{HillClimb}{$\mathcal{C}^{(0)}$, $\vec{\theta}^{(0)}$}
            \State $E_0 \gets \mathcal{E}(\mathcal{C}^{(0)}(\vec{\theta}^{(0)}))$
            \For{$k \gets 1\,..\,k_\text{max}$}
                \State $E_k \gets E_{k-1}$
                \ForAll{$M \in \textproc{GenerateMoves}(\mathcal{C}^{(k-1)})$}
                    \State $\tilde{\mathcal{C}}, \tilde{\vec{\theta}} \gets \textproc{ApplyMove}(\mathcal{C}^{(k-1)}, \vec{\theta}^{(k-1)})$
                    \State $\tilde{\vec{\theta}} \gets \textproc{OptimiseParameters}(\tilde{\mathcal{C}}, \tilde{\vec{\theta}})$
                    \If{$\mathcal{E}(\tilde{\mathcal{C}}(\tilde{\vec{\theta}})) < E_k$}
                        \State $\mathcal{C}^{(k)} \gets \tilde{\mathcal{C}}$,\quad $\vec{\theta}^{(k)} \gets \tilde{\vec{\theta}}$,\quad $E_k \gets \mathcal{E}(\tilde{\mathcal{C}}(\tilde{\vec{\theta}}))$
                    \EndIf
                \EndFor
                \If{$E_k < E_\text{conv}$}
                    \State \Return $\textproc{Prune}(\mathcal{C}^{(k)}, \vec{\theta}^{(k)}, \{0, \ldots, N_\mathrm{gates}(\mathcal{C}^{(k)})\})$
                \EndIf
                \If{$E_k = E_{k-1}$}
                    \State \Return \textbf{fail} \Comment{No more improvement}
                \EndIf
            \EndFor
            \State \Return \textbf{fail} \Comment{No convergence at iteration limit}
        \EndFunction
    \end{algorithmic}
    \caption{A simple variant of hill climbing for circuit synthesis.}
    \label{algo:hillclimb}
\end{algorithm}

\clearpage

\end{document}